\documentclass[twocolumn,showpacs,floatfix,amsmath,amssymb,prb,superscriptaddress]{revtex4}
\usepackage[dvips]{graphicx} 
\usepackage{latexsym} 
\usepackage{graphicx} 
\usepackage{times,psfrag,subfigure} 
\usepackage{amsmath} 
\usepackage{dcolumn} 
\usepackage{latexsym,amsmath,amssymb,bm,euscript} 
\def\bfx{{\bf x}} 
\def\barpsi{\overline{\psi}} 
\def\CsCuCl{$\text{Cs}_2\text{CuCl}_4$} 
 
\def\DM{\text{Dzyaloshinskii-Moriya}}

\begin{document}

\title{Theory of the algebraic vortex liquid in an anisotropic  
spin-1/2 triangular antiferromagnet} 
\author{Jason Alicea} 
\affiliation{Physics Department, University of California, Santa 
Barbara, CA 93106} 
\author{Olexei I. Motrunich} 
\affiliation{Kavli Institute for Theoretical Physics, University of 
California, Santa Barbara, CA 93106} 
\author{Matthew P. A. Fisher} 
\affiliation{Kavli Institute for Theoretical Physics, University of 
California, Santa 
Barbara, CA 93106} 
 
\date{\today} 
 
\begin{abstract} 
 
We explore spin-1/2 triangular antiferromagnets with both easy-plane and 
lattice exchange anisotropies by employing a dual vortex mapping followed by a 
fermionization of the vortices.  Over a broad range of exchange 
anisotropy, this approach leads naturally to  
a ``critical'' spin liquid---the algebraic vortex 
liquid---which appears to be distinct from other known spin liquids. 
We present a detailed characterization of this state, which  
is described in terms of non-compact QED3 with an emergent SU(4) 
symmetry.  Descendant phases of the algebraic vortex liquid are also 
explored, which include the Kalmeyer-Laughlin spin liquid, a  
variety of magnetically ordered states 
such as the well known coplanar spiral state, and supersolids. 
In the range of exchange 
anisotropy where the ``square lattice'' Neel ground state arises, we 
demonstrate that anomalous ``roton'' minima in the excitation spectrum  
recently reported in series expansions can be accounted for within  
our approach.   
 
\end{abstract} 
 
\maketitle 
 
\section{Introduction} 
\label{Intro}

A fundamental theoretical challenge in strongly 
correlated systems lies in understanding the behavior of frustrated 
quantum magnets, whose properties often bear little resemblance to 
those of their classical counterparts. 
In the most exotic scenario, quantum fluctuations are sufficiently 
strong to disorder the system even at zero temperature, and a spin 
liquid ground state emerges.   
Historically, Anderson originally suggested that the spin-1/2 
Heisenberg triangular antiferromagnet may realize such a  
quantum-disordered ground state.\cite{RVB} 
It is now  
recognized that with only nearest-neighbor exchange the true  
ground state on the triangular lattice is the magnetically ordered 
$\sqrt{3}\times\sqrt{3}$ phase, though the order is  
significantly diminished relative 
to the classical state.\cite{Huse, Sorella}  
It is conceivable, then, that a spin liquid may arise with not too 
drastic perturbations to the model, and the triangular lattice has 
thus remained a prominent setting in the search for two-dimensional 
spin liquids.   
 
Recent experiments on the spin-1/2 anisotropic triangular 
antiferromagnet \CsCuCl\ stimulated renewed interest in possible 
spin-liquid phases proximate to the nearest-neighbor Heisenberg 
model.\cite{ColdeaShort,ColdeaLong}    
This material is accurately modeled by an anisotropic 
Heisenberg Hamiltonian supplemented by a weak  
\DM~interaction.\cite{ColdeaHamiltonian}  Although long-range spiral order  
develops at temperatures 
$T\lesssim 0.62$K, the dynamical structure factor measured via 
neutron scattering exhibits ``critical'' power laws at intermediate 
energies, both in the ordered phase and in a range of temperatures 
above $T_N$.  This unusual power law behavior in the excitation spectrum  
is highly suggestive of spinon deconfinement  
that is characteristic of spin liquids.    
 
A variety of theoretical approaches have been employed to capture spin 
liquids on the triangular lattice. 
Kalmeyer and Laughlin exploited a mapping between the spin-1/2 
Heisenberg model and hard-core bosons in a magnetic field to obtain a 
``chiral'' spin liquid which breaks time-reversal 
symmetry.\cite{KLshort,KLlong}  Their arguments were subsequently 
reformulated by Yang \emph{et al}.,\cite{Yang} who arrived at the 
chiral spin liquid by fermionizing the spins using Chern-Simons flux 
attachment\cite{JordanWigner} and expanding around a ``flux-smeared'' 
mean-field state.  Using a slave-boson representation of the spin 
operators, Sachdev explored an Sp($N$) generalization of the 
Heisenberg model, and in the large-$N$ limit obtained a Z$_2$ spin 
liquid ground state, which breaks no symmetries.\cite{SachdevZ2}   
The Z$_2$ spin liquid was later 
realized microscopically in a quantum dimer model on the triangular  
lattice.\cite{Moessner}  Finally, a large class of spin liquids was 
studied by Zhou and Wen using a slave fermion 
representation of the spins.\cite{ZhouWen}  
Whereas excitations in both the chiral and Z$_2$ spin liquids are 
gapped, the slave fermion mean-field approach can give rise to so called 
``algebraic spin liquids'', which admit \emph{gapless}  
spin excitations and power-law spin correlations.   
 
In this paper we pursue an alternate approach to the spin-1/2 
triangular antiferromagnet, and use vortex duality to attack the problem  
coming from the easy-plane regime.  Duality has been a 
powerful tool for exploring unconventional phases such as  
valence bond solids and spin liquids in quantum spin  
systems,\cite{ReadSachdev, SachdevPark, Lannert, DECCP, Z2gaugetheory}  
as well as complex charge-ordered states  
in bosonic systems.\cite{duality, Tesanovic1, Tesanovic2, Burkov1, 
  Burkov2, Burkov} 
The main difficulty here is that vortices are at  
finite density, 
which is familiar from dual approaches to the fractional  
quantum Hall problem. 
As an initial step towards applying duality to frustrated spin systems, 
in Ref.\ \onlinecite{spin1} we examined integer-spin triangular  
antiferromagnets with easy-plane symmetry from the vortex perspective. 
By \emph{fermionizing} the vortices using Chern-Simons flux 
attachment, it was shown that an effective low-energy dual formulation 
can be derived, which was argued to reproduce the physics of a more 
direct Landau-Ginzburg-Wilson analysis of the spin model. 
While this approach is reminiscent of the spin fermionization adopted 
by Yang \emph{et al}.,\cite{Yang} we emphasize that alternatively 
fermionizing vortices is advantageous because the vortices interact 
logarithmically, which allows for a more controllable treatment of 
Chern-Simons gauge fluctuations. 
 
Here we extend the fermionized vortex approach to the spin-1/2  
triangular antiferromagnet with 
easy-plane symmetry and anisotropic nearest-neighbor exchanges $J$ and 
$J'$ as shown in Fig.\ \ref{lattice}.  This formalism allows us 
to explore the phase diagram of the spin 
model in a setting where a more conventional Landau analysis of the 
spin model is \emph{not} accessible due to Berry phases.   
Remarkably, over a broad range of anisotropy $J'/J \lesssim 1.4$ this 
approach leads naturally, with the simplest flux-smeared mean-field 
starting point, to a novel ``critical'' spin liquid that we will refer to 
as the algebraic vortex liquid.  This state was introduced earlier  
and applied to 
\CsCuCl~in a short letter, Ref.\ \onlinecite{AVLshort}, and is  
characterized in detail 
here.  Schematically, vortices form a critical state with four 
Dirac nodes, and interact via a fluctuating gauge field 
representing the original boson current fluctuations. 
Already on the mean-field level, the gapless character of the vortex  
state implies power-law $S^z$ and $S^\pm$ spin correlations at specific 
wave vectors shown in Figs.\ \ref{BilinearQs} and \ref{MonopoleQs},  
respectively. 
Such momenta for low-energy excitations in the spin system are  
determined by short-distance physics in the frustrated magnet, 
and we propose that this physics is well-captured in the vortex  
treatment. 
 
Going beyond a mean-field analysis, we argue that the algebraic  
vortex liquid is 
described by QED3 with an emergent global SU(4) flavor symmetry, which 
has further implications for the dynamical spin correlations. 
In particular, as a consequence of the SU(4) symmetry the in-plane spin 
structure factor exhibits enhanced universal power law correlations 
with the \emph{same exponent} at several momenta in the Brillouin 
zone: the spiral ordering wave vectors $\pm {\bf Q}$ and momenta ${\bf 
  K}_{1,2,3}$ at the midpoints of the Brillouin zone edges (see Fig.\ 
\ref{MonopoleQs}).  The out-of-plane spin structure factor meanwhile has 
enhanced correlations only at the spiral ordering wave vectors $\pm 
{\bf Q}$.  These nontrivial properties distinguish the algebraic 
vortex liquid from other known spin liquids.   
Interestingly, the prominence of momenta ${\bf K}_{1,2,3}$ in the 
theory appears to be consistent with recent series  
expansion studies of the Heisenberg triangular 
antiferromagnet\cite{ZhengSeriesExpansion}, which observe  
excitation energies at these wave vectors which are dramatically  
reduced relative to linear spin wave theory.   
Moreover, the prediction of 
``active'' momenta $\pm {\bf Q}$ and ${\bf K}_{1,2}$ in the anisotropic 
system seems to capture the neutron scattering data for  
\CsCuCl.\cite{ColdeaShort,ColdeaLong}

The phase diagram in the vicinity of the algebraic vortex liquid is 
also explored, and found to be rather rich.  Nearby phases include  
the Kalmeyer-Laughlin chiral spin  
liquid, numerous magnetically ordered states including the coplanar  
spiral state, and variants of supersolids discussed  
recently.\cite{RogerSS, DamleSS, TroyerSS, ProkofevSS}

In the range of anisotropy $J'/J \gtrsim 1.4$, our treatment captures 
the ``square-lattice'' Neel ordered state, which is the expected 
ground state in this regime\cite{Zheng99}.  Here, we demonstrate  
in a particularly clear setting that 
anomalous ``roton'' minima in the excitation spectrum observed by  
series expansion studies can indeed  
be accounted for as low energy vortex-antivortex 
excitations.\cite{Zheng99,ZhengSeriesExpansion}  We 
further predict that these low-energy rotons may have still more  
dramatic effects in the easy-plane regime, which would be useful 
to explore using series expansions.

The paper is organized as follows.  The spin model 
and the dual vortex mapping are introduced in Sec.\ II.  In Sec.\ III 
the fermionized vortex theory is developed.  We first discuss the ``roton'' 
excitations in the Neel phase arising when $J'/J \gtrsim 1.4$. 
We then obtain a low-energy effective theory for $J'/J \lesssim 1.4$ which 
contains a description of the algebraic vortex liquid.   
Sec.\ IV focuses on the 
properties of the algebraic vortex liquid, including its stability, 
symmetries, and dynamic spin correlations.  The proximate phases of 
the algebraic vortex liquid are explored in Sec.\ V, and we conclude 
with a discussion in Sec.\ VI.

\section{Model} 
\label{Model}

\subsection{Easy-plane Spin Model} 
\label{SpinModel} 
\begin{figure}  
  \begin{center}  
    {\resizebox{7cm}{!}{\includegraphics{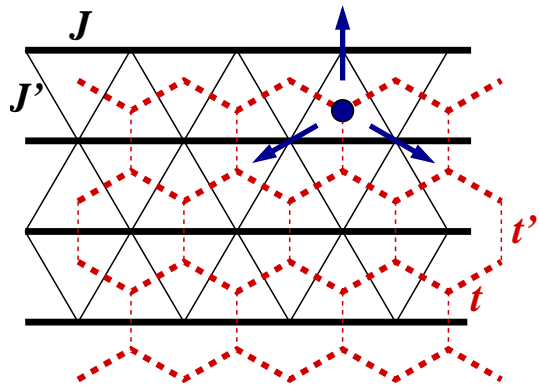}}}  
  \end{center}  
  \caption{Triangular lattice and the dual honeycomb on which vortices 
    reside.  Spins shown illustrate a vortex.   
    The spin exchange and vortex hopping amplitudes are generally 
    anisotropic, with $J'/J \sim t'/t$.  }  
  \label{lattice}  
\end{figure}

We begin by considering an easy-plane, anisotropic spin-1/2 triangular 
antiferromagnet modeled by an XXZ Hamiltonian with nearest-neighbor exchange, 
\begin{equation} 
  H_0 = \frac{1}{2}\sum_{\langle{\bf r r'}\rangle}J_{\bf r r'} 
  [S_{\bf r}^+ S_{\bf r'}^- + \text{H.c.}] +  
  \sum_{\langle{\bf r r'}\rangle}J^z_{\bf r r'} S^z_{\bf r} S^z_{\bf r'}, 
  \label{H} 
\end{equation} 
where $S_{\bf r}^\pm = S^x_{\bf r} \pm i S^y_{\bf r}$ are the usual spin 
raising and lowering operators.  As illustrated in 
Fig.\ \ref{lattice}, we take the 
in-plane exchange energy to be $J_{\bf r r'} = J$ along bold 
horizontal links and $J_{\bf r r'} = J'$ along diagonal links of 
the triangular lattice.  The out-of-plane exchange is defined to be  
$J^z_{\bf r r'} \equiv \gamma J_{\bf r r'}$, with $0<\gamma<1$  
to satisfy the easy-plane condition.   
 
It is convenient to work with the easy-plane spin model  
recast in terms of quantum rotors by introducing 
an integer-valued boson number $n_{\bf r}$ and its conjugate 
phase $\varphi_{\bf r}$.  Upon identifying $S^z_{\bf 
r} \rightarrow n_{\bf r}-1/2$ and $S^+_{\bf r} \rightarrow 
e^{i\varphi_{\bf r}}$, the Hamiltonian reads 
\begin{eqnarray} 
  H_{0} &=& \sum_{\langle{\bf r r'}\rangle}J_{\bf r r'} 
  \cos(\varphi_{\bf r}-\varphi_{\bf r'}) +  
  U\sum_{\bf r}(n_{\bf r}-1/2)^2 
  \nonumber \\ 
  &+& \sum_{\langle{\bf r r'}\rangle}J^z_{\bf r r'}(n_{\bf r}-1/2) 
  (n_{\bf r'}-1/2). 
  \label{Hrotor} 
\end{eqnarray} 
The $U$ term above energetically enforces the 
constraint of having either 0 or 1 boson per site as required for 
modeling a spin-1/2 system.   
 
The XXZ Hamiltonian $H_0$ respects a number of internal 
and discrete lattice symmetries which we now enumerate.   
The model exhibits U(1) 
spin symmetry  and is invariant under time reversal ${\mathcal T}$ and 
a ``particle-hole'' transformation ${\mathcal C}$ which sends  
$(S^x,S^y,S^z)\to(S^x,-S^y,-S^z)$.  Under these 
operations, the rotor fields transform as follows: 
\begin{eqnarray} 
  U(1) &:& n \to n , \quad 
      e^{i\varphi} \to e^{i(\varphi+\alpha)}  
  \\ 
  {\mathcal C} &:& n \to 1-n , \quad e^{i\varphi} \to 
      e^{-i\varphi} 
  \\ 
  {\mathcal T} &:& n \to 1-n , \quad e^{i\varphi} \to 
      -e^{-i\varphi}, 
\end{eqnarray} 
where $\alpha$ is a constant and ${\mathcal T}$ is an antiunitary 
operation which sends $i\to -i$.  The model also preserves  
translations $T_{\bf \delta r}$ by triangular lattice vectors  
${\bf \delta r}$, as well as $x$-reflections ${\mathcal R}_{x}$ and 
inversions (\emph{i.e.}, $\pi$ rotations) $R_\pi$ about a triangular  
lattice site.  Rather 
than considering the $x$-reflections ${\mathcal R}_{x}$, it will be  
useful for subsequent developments to work with a modified 
antiunitary reflection $\tilde {\mathcal R}_{x} \equiv 
{\mathcal R}_{x} {\mathcal C} {\mathcal T}$.  The latter operations 
transform the rotor fields as 
\begin{equation} 
  \tilde {\mathcal R}_{x}: n_{\bf r} \to n_{\bf r'} , \quad 
  e^{i\varphi_{\bf r}} \to -e^{i\varphi_{\bf r'}}, 
\end{equation} 
where ${\bf r'}$ is an appropriately reflected coordinate.  In the 
isotropic limit $J = J'$, the XXZ Hamiltonian additionally preserves 
$\pi/3$ rotations $R_{\pi/3}$ about a triangular lattice site; 
$R_{\pi}$ is then no longer an independent symmetry since  
$R_{\pi/3}^3 = R_{\pi}$.

In this paper we are interested in exploring the phase diagram 
accessible with the above XXZ spin model as a starting point.   
In particular, as the low spin and geometric frustration strongly suppress the 
tendency to magnetically order, it is natural to ask whether  
spin-liquid phases can be realized with not too drastic perturbations 
to the model. 
To this end, we would like to derive an effective theory 
that governs the low-energy behavior of the spin system. 
This is, for instance, readily achieved for the integer-spin analogue of Eq.\ 
(\ref{H}), and the phase diagram can be explored 
within a standard Landau analysis.  However, for the spin-1/2 system 
studied here a similar direct analysis of the spin model  
is hindered by the presence of Berry phases.  Consequently, obtaining  
a low-energy  
theory is largely intractable in this formulation.   
 
To proceed we utilize an alternative dual approach, 
introduced in the context of integer-spin systems in Ref.\ 
\onlinecite{spin1}, wherein one considers a reformulation 
of the problem in terms of \emph{fermionized vortices}.  In this 
framework, the basic degrees of freedom one works with are  
vortices---topological defects in which the phases  
$\varphi_{\bf r}$ of the spins wind by $2\pi$ around a triangular  
plaquette as in Fig.\ \ref{lattice}---rather than the spins themselves. 
Although the vortices as defined are bosonic, it 
will prove extremely useful to fermionize them in a manner familiar 
from the fractional quantum Hall effect via Chern-Simons flux 
attachment.  Doing so enables us to obtain a low-energy dual 
theory, which as we will demonstrate leads naturally to a novel 
``critical'' spin-liquid phase, the algebraic vortex 
liquid.

\subsection{Dual Vortex Mapping} 
\label{DualTheory} 
 
We proceed now to the dual vortex theory.  We forgo the details of 
the duality mapping as these are provided in Sec.\ III of  
Ref.\ \onlinecite{spin1} 
in a very similar setting, and instead emphasize the important physical 
aspects of the dual theory.  Implementing the duality 
transformation on the quantum rotor Hamiltonian  
Eq.\ (\ref{Hrotor}),\cite{duality}  
one obtains a theory of bosonic vortices with ``electromagnetic'' interactions 
hopping among sites of the dual honeycomb lattice depicted by the 
dashed lines in Fig.\ \ref{lattice}.    
The vortices interact via a ``vector  
potential'' $a_{\bf x x'}\in \mathbb{R}$ and a conjugate  
``electric field'' $e_{\bf x x'}$ which reside on honeycomb links and 
mediate a logarithmic vortex repulsion.   Here ${\bf 
  x}, {\bf x'}$ denote nearest-neighbor honeycomb sites. 
(Throughout, we distinguish sites of the honeycomb and triangular 
lattices by the labels ``${\bf x}$'' and ``${\bf r}$,'' respectively.) 
These dual gauge fields satisfy the commutation relation $[e_{\bf x 
    x'},a_{\bf x x'}] = i$ and commute on different links. 
In-plane spin components are encoded in this formulation through  
the ``electric field'' and the vortices. 
The $S^z$ component of spin meanwhile appears as a dual 
``magnetic flux,'' 
\begin{equation} 
  S^z_{\bf r} \sim \frac{1}{2\pi} (\Delta\times a)_{\bf r}, 
  \label{Sz} 
\end{equation} 
where $(\Delta\times a)_{\bf r}$ signifies a lattice curl of $a_{\bf x x'}$ 
around the hexagon encircling site ${\bf r}$ of the triangular 
lattice.  Although $a_{\bf x x'}$ roams over the real numbers,  
the desired half-integer values of $S^z$ in Eq.\ (\ref{Sz}) 
are imposed energetically in the dual theory.

In terms of a vortex number operator $N_{\bf x}$ and vortex creation 
operator $e^{i\theta_{\bf x}}$, the dual vortex Hamiltonian can be  
expressed as 
\begin{eqnarray} 
  {\mathcal H}_{\text{dual}} = {\mathcal H}_a 
  - \sum_{\langle {\bf x x'} \rangle}2 t_{\bf x x'}   
  \cos(\theta_{\bf x}-\theta_{\bf x'} -a_{\bf x x'}-a^0_{\bf x x'}) 
  \label{Hdual}  
\end{eqnarray} 
together with a Gauss's law constraint for the ``electric field'', 
\begin{equation} 
  (\Delta\cdot e)_{\bf x} = N_{\bf x}-1/2. 
\end{equation} 
Here $a^0_{\bf x x'}$ is a static gauge field satisfying  
$(\Delta \times a^0)_{\bf r} = \pi$, and $(\Delta\cdot e)_{\bf x}$ 
denotes a lattice divergence of $e_{\bf x x'}$ at site ${\bf x}$.   
Moreover, ${\mathcal H}_a$ describes the gauge field dynamics, 
\begin{eqnarray} 
  {\mathcal H}_a &=& \sum_{\langle {\bf x x'} \rangle}   
  {\mathcal J}_{\bf x x'} e_{\bf x x'}^2  
  + {\mathcal U} \sum_{\bf r}(\Delta \times a)_{\bf r}^2  
  \nonumber \\ 
  &+& \sum_{\langle {\bf r r'} \rangle} {\mathcal J}^z_{\bf r r'} 
  (\Delta \times a)_{\bf r}(\Delta \times a)_{\bf r'}, 
\end{eqnarray}   
with ${\mathcal U} = U/(2\pi)^2$, ${\mathcal J}^z_{\bf r r'} = 
J^z_{\bf r r'}/(2\pi)^2$, and ${\mathcal J}_{\bf x x'} = 
2 \pi^2 J'$ on the bold zigzag honeycomb links in Fig.\ 
\ref{lattice} while ${\mathcal J}_{\bf x x'} = 2 \pi^2 J$ on  
vertical honeycomb links.   
 
The cosine term in Eq.\ (\ref{Hdual}) describes nearest-neighbor  
vortex hopping in an average background of $\pi$ flux per hexagon.   
This background ``magnetic flux'' is provided by the static gauge field  
$a^0_{\bf x x'}$ and arises because $S^z$ is half-integer valued in  
the original spin model.  (The average background flux for an integer spin 
system, in contrast, is trivial.\cite{spin1}) 
The hopping amplitudes $t_{\bf x x'}$ are chosen to be anisotropic to 
reflect the spin exchange anisotropy.   
In particular, as illustrated in Fig.\ \ref{lattice} we take  
$t_{\bf x x'} = t$ on the bold zigzag links of the honeycomb  
and $t_{\bf x x'} = t'$ on vertical links, with $t'/t \sim J'/J$ since 
vortices hop more easily across weak spin links than strong spin 
links.   
 
An important feature of the dual theory is that with our conventions\cite{spin1}  
the bosonic vortices are at \emph{half-filling}, which is a direct  
consequence of the underlying frustration in the spin model.  For 
example, in the classical $\sqrt{3}\times\sqrt{3}$ spin-ordered 
state we define the vortex number to be one on ``up'' triangles and zero on 
``down'' triangles (or vice versa, depending on the chirality).   
The half-filling of the 
vortices becomes manifest upon reexpressing the dual Hamiltonian in 
terms of an \emph{unconstrained} electric field as follows, 
\begin{eqnarray} 
  \tilde {\mathcal H}_{\rm dual} = {\mathcal H}_{\rm dual} 
  +\sum_{\bf x x'} (N_{\bf x}-1/2)V_{\bf x x'}(N_{\bf x'}-1/2), 
  \label{Hdual2} 
\end{eqnarray} 
where $V_{\bf x x'}$ encodes the logarithmic vortex repulsion. 
Equation (\ref{Hdual2}) clearly exhibits a vortex particle-hole 
symmetry.   
 
The transformation properties of the dual fields under the discrete microscopic 
symmetries can be straightforwardly deduced as discussed in Ref.\ 
\onlinecite{spin1}.  These are summarized in Table \ref{tab:dual}. 
The continuous U(1) spin symmetry, which reflects conservation  
of $S^z$, is 
not directly manifest in this formulation and is instead replaced by 
a conservation of dual gauge flux, $(\Delta\times a)$. 
Additionally, the dual Hamiltonian has a U(1) gauge redundancy, being 
invariant under $(a_{\bf x x'}+a^0_{\bf x x'}) \to (a_{\bf x 
  x'}+a^0_{\bf x x'}) + \Lambda_{\bf x} 
-\Lambda_{\bf x'}$ and $\theta_{\bf x} \to \theta_{\bf x} + 
\Lambda_{\bf x}$ for arbitrary $\Lambda_{\bf x} \in {\mathbb R}$.   
 
\begin{table}  
\caption{\label{tab:dual} Transformation properties of fields  
in the dual bosonic-vortex formulation under the discrete  
microscopic symmetries.  The lattice coordinates,  
which also transform appropriately under the lattice symmetries,  
have been suppressed on all fields for brevity.   }  
\begin{ruledtabular}  
\begin{tabular}{c | c | c }  
  $T_{{\bf \delta r}}$, $R_\pi$, $R_{\pi/3}$ (isotropic limit)   
  &  ${\mathcal C}$  
  & $\tilde {\mathcal R}_{x}$, ${\mathcal T}$ \\  
  \hline  
  $a^0\rightarrow a^0$   
  & $a^0\rightarrow-a^0$  
  & $a^0\rightarrow -a^0$ \\  
  \hline  
  $a\rightarrow a$   
  & $a\rightarrow-a$  
  & $a\rightarrow -a$ \\  
  \hline  
  $ e \rightarrow e$  
  & $e \rightarrow -e $  
  & $ e\rightarrow e$ \\  
  \hline  
  $\theta\rightarrow\theta$   
  & $\theta\rightarrow-\theta$  
  & $\theta\rightarrow-\theta$ \\  
  \hline  
  $N\rightarrow N$   
  & $N\rightarrow 1-N$   
  & $N\rightarrow N$ 
\end{tabular}  
\end{ruledtabular}  
\end{table}

\section{Fermionized-Vortex Formulation} 
\label{FermVortFormulation}

Due to the finite vortex density together with the strong vortex 
interactions, the dual theory as it stands appears as intractable as 
the original spin model.  There is, however, an important distinction  
between the dual vortex formulation and the original hard-core  
boson representation of the spin model that we can exploit. 
In the dual theory the vortices move in the presence of a 
dynamical gauge field which encodes the motion of the  
hard-core bosons.  Thus, the dual vortex theory is in some sense 
a two-fluid model that describes simultaneously both the dynamics of the 
hard-core bosons and the vortices.   As such, it is possible to 
imagine a vortex moving together with a cloud of dual gauge  
flux, $(\Delta\times a)$, 
which can in effect modify the statistics of the vortex-flux composite. 
Indeed, if the flux has strength $\pm 2\pi$, the composite particle will 
behave like a fermion due to the Aharonov-Bohm phase 
acquired from the dual flux under an exchange process. 
What we imagine is that the motion of the vortices and the dual flux 
can become dynamically correlated in such a fashion to be well 
represented (on intermediate length scales) by the dynamics of fermionic  
vortex-flux composites moving in the presence of the remaining  
dynamical gauge flux. 
 
This physical picture can be implemented by splitting the gauge flux 
into two pieces, $a \rightarrow a + A$, and attaching $2\pi$ flux of 
$(\Delta\times A)$ to the vortices with the help of a Chern-Simons 
term for $A$.  An unfortunate but apparently unavoidable feature 
of this Chern-Simons approach is that we have to choose the sign 
of the attached flux, say, $+ 2\pi$ rather than $-2\pi$. 
One could contemplate an alternate formulation wherein the 
sign of the attached Chern-Simons flux is itself a dynamically  
fluctuating field, but we do not attempt to do so in this paper.

Before proceeding to the details, we pause to comment on the usefulness 
and limitations of this approach.   
Working with fermionized vortices is  
expected to be legitimate for describing physics in regimes 
where the vortex exchange statistics is unimportant.   
Consider, for instance, ``insulating'' phases of the vortices, 
examples of which include vortex crystals and ``valence bond solids''. 
Such phases were explored in integer-spin systems in  
Ref.\ \onlinecite{spin1}, and shown to  
correspond to magnetically ordered spin states.   
At the lowest energy scales, vortex density fluctuations are entirely 
frozen out,  
rendering their exchange statistics unimportant; 
whether the vortices are treated as bosonic or fermionic is presumably 
inconsequential.   
 
On the other hand, once approximations are made  
to derive a low energy effective theory as we will do below, 
the vortex fermionization approach  
is expected 
to be least reliable  
when describing ``vortex condensates''.
(Such vortex condensates correspond to paramagnetic states of the 
original spin system.)  It is intuitively clear that 
trying to mimic the physics of Bose condensation will 
be challenging with fermionic fields, although this 
was the approach taken to describe anyon superconductivity  
by a number of authors some years back. 
 
Here, we will be most interested in employing the fermionized 
vortex approach to access the ``critical" algebraic vortex liquid.   
As we shall see, although the 
vortices are mobile in this phase, due to their 
long-range interactions vortex density 
fluctuations will be so strongly suppressed that  
the Chern-Simons flux attachment will 
be ineffective at modifying the behavior on long length scales. 
We will argue that the asymptotic low-energy physics of the 
algebraic vortex liquid is described by fermionic vortices 
minimally coupled to a gauge field mediating a long-range interaction 
(with Maxwell but no Chern-Simons term).  This theory is usually 
referred to as QED3.

\subsection{Fermionization} 
\label{Fermionization} 
 
Formally, fermionization can be implemented by treating the vortices 
as hard-core bosons, replacing $e^{i\theta_{\bf x}}\to b^\dagger_{\bf 
x}$ and $N_{\bf x} \to b^\dagger_{\bf x} b_{\bf x} = 0,1$, followed by   
a 2D Jordan-Wigner transformation,\cite{JordanWigner} 
\begin{eqnarray} 
  b_\bfx^\dagger &=& d_\bfx^\dagger  
  \exp[i \sum_{\bfx' \neq \bfx} \arg(\bfx, \bfx') N_{\bfx'} ],  
  \label{JW}  
  \\  
  N_\bfx &=& b_\bfx^\dagger b_\bfx = d_\bfx^\dagger d_\bfx.  
\end{eqnarray}  
Here ${\rm arg}({\bf x},{\bf x'})$ denotes an angle formed by the vector 
${\bf x} - {\bf x'}$ with respect to an arbitrary fixed axis. 
 
The dual fermionized-vortex Hamiltonian takes the form 
\begin{eqnarray} 
  {\mathcal H}_{\text{dual}} &=&  
  - \sum_{\langle {{\bf x}_1 {\bf x}_2} \rangle} t_{{\bf x}_1 {\bf x}_2}   
  [d^\dagger_{{\bf x}_1} d_{{\bf x}_2} e^{-i(a_{{\bf x}_1 {\bf x}_2} 
  +a^0_{{\bf x}_1 {\bf x}_2}+A_{{\bf x}_1 {\bf x}_2})}  
  \nonumber \\ 
  &+& {\rm H.c.}] + {\mathcal H}_a, 
  \label{HdualFerm}  
\end{eqnarray} 
where we have introduced a Chern-Simons field 
\begin{equation} 
  A_{{\bf x}_1 {\bf x}_2} = \sum_{{\bf x'} \neq {\bf x}_1, {\bf x}_2}  
  [{\arg}({\bf x}_2,{\bf x'}) - {\arg}({\bf x}_1,{\bf 
  x'})]N_{\bf x'} 
  \label{CS} 
\end{equation} 
which in Eq.\ (\ref{HdualFerm}) resides on honeycomb links.   
Although we have included only nearest-neighbor hopping in the dual vortex 
Hamiltonian, one could also generically allow for small further-neighbor 
hopping terms allowed by symmetry. 
Upon fermionization, such terms similarly involve fermions coupled  
to a Chern-Simons field 
defined as in Eq.\ (\ref{CS}), but with ${\bf x}_{1}$ and ${\bf x}_2$  
further-neighbor sites.   
 
The transformation properties of the fermions and the Chern-Simons 
field can be deduced by examining Eqs.\ (\ref{JW}) and 
(\ref{CS}).  Table \ref{tab:d} summarizes the symmetry properties of 
all fields in this representation.  According to Eq.\ (\ref{JW}), 
particle-hole symmetry sends  
$d_{\bf x} \rightarrow d_{\bf x}^\dagger e^{i\gamma_{\bf x}}$, where for  
nearest-neighbor honeycomb sites ${\bf x}_{1,2}$ the acquired  
phases satisfy 
$\gamma_{{\bf x}_1} - \gamma_{{\bf x}_2} = \pi-2\langle A_{{\bf 
    x}_1{\bf x}_2}\rangle$. 
Here $\langle A_{{\bf x}_1{\bf x}_2}\rangle$ denotes the mean-field 
value of the Chern-Simons field with $\langle N_{\bf x}\rangle = 
1/2$ appropriate for half-filled fermions.  Since the Chern-Simons 
flux through a given hexagonal plaquette averages to $2\pi$, which is 
equivalent to zero flux, we take $\langle A\rangle = 0$ on 
nearest-neighbor links.  Hence, in the table we 
implement particle-hole symmetry by transforming $d_{\bf x} 
\rightarrow (-1)^j d_{\bf x}^\dagger$, where $j = 1,2$ labels one of 
the two sublattices of the honeycomb. 
Note also that as discussed in Ref.\ 
\onlinecite{spin1} time reversal acts nonlocally on the fermions, and 
consequently we do not know how to faithfully realize this symmetry in 
the continuum theory derived in the next subsection.  In the last 
column of Table \ref{tab:d} we provide a modified time 
reversal, ${\mathcal T}_{\rm ferm}$,  
which acts locally on the fermion fields and corresponds to naive time  
reversal for fermions on a lattice.

\begin{table*}  
\caption{\label{tab:d} Transformation properties of the fields  
in the dual fermionized-vortex representation.  Symmetry properties 
of $a^0, a, e$ are the same as in Table~\ref{tab:dual}. 
In the ${\mathcal C}$ column, $j=1,2$ labels one of the two triangular 
sublattices of the honeycomb, and $\langle A \rangle$ refers to the 
mean-field value of $A_{\bf x x'}$ with $\langle N_{\bf x} \rangle = 1/2$.   
The additional column ${\mathcal T}_{\rm ferm}$ 
corresponds to the naive time reversal for the lattice fermions and 
is \emph{not} a symmetry of the vortex Hamiltonian.}  
\begin{ruledtabular}  
\begin{tabular}{c | c | c | c | c || c}  
  & $T_{\delta {\bf r}}$, $R_\pi$, $R_{\pi/3}$ (isotropic limit)   
  & $\tilde{\mathcal R}_{x}$  
  & ${\mathcal C}$  
  & ${\mathcal T}$  
  & ${\mathcal T_{\rm ferm}}$ \\    
  \hline  
  $a^0\to$ 
  & $a^0$   
  & $-a^0$  
  & $-a^0$  
  & $-a^0$  
  & $-a^0$ \\  
  \hline  
  $a\to$ 
  & $a$   
  & $-a$  
  & $-a$  
  & $-a$  
  & $-a$ \\  
  \hline  
  $e \to$  
  &$e$  
  & $e$  
  & $-e$  
  & $e$  
  & $e$ \\  
  \hline  
  $d_\bfx \to$ 
  & $d$  
  & $d$ 
  & $(-1)^j d^\dagger$  
  & $d_\bfx e^{-2 i \sum_{\bfx' \neq \bfx} \arg(\bfx, \bfx') N_{\bfx'} }$ 
  & $d$ 
  \\  
  \hline  
  $A \to $ 
  & $A$  
  & $-A$   
  & $2 \langle A \rangle-A$  
  & $A$ \\  
\end{tabular}  
\end{ruledtabular}  
\end{table*}  
  
\subsection{Vortex Mean Field and Low-energy Theory} 
\label{LowEnergyTheory} 
 
One advantage of working with fermionized vortices is that there is then a 
natural route to a low-energy effective theory.  Namely, we start by  
considering a 
non-interacting ``flux-smeared'' mean-field state, ignoring fluctuations in the 
Chern-Simons and ``electromagnetic'' gauge fields and replacing the flux 
by an average background.  Since the vortices are at half-filling, the 
Chern-Simons flux through each hexagon averages to $2\pi$, which is 
equivalent to zero flux on a lattice.  Thus, the ``flux-smeared'' mean-field 
Hamiltonian describes free fermionic vortices hopping on the  
honeycomb in a background of $\pi$ flux (due to $a^0_{\bf x x'}$): 
\begin{equation} 
  \mathcal{H}_{\rm MF} = -\sum_{\langle {\bf x x'} \rangle} t_{\bf 
  x x'} (d^\dagger_{\bf x} d_{\bf x'} e^{-i a^0_{\bf x x'}} + 
  \text{H.c.}). 
  \label{MeanFieldH} 
\end{equation} 
 
\begin{figure}  
  \begin{center}  
    {\resizebox{5.5cm}{!}{\includegraphics{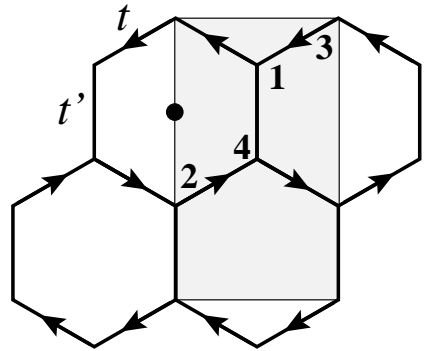}}}  
  \end{center}  
  \caption{Four-site unit cell chosen for the honeycomb.  With our 
  gauge choice, the static gauge field $a^0_{\bf x x'}$ is 
  zero on the vertical links, 
  while $a^0_{\bf x x'}=\pi/4$ on the zigzag links directed 
  along the arrows.  The filled circle indicates our origin, 
  which coincides with a triangular lattice site.}  
  \label{UnitCell}  
\end{figure}

\begin{figure}  
  \begin{center}  
    {\resizebox{6cm}{!}{\includegraphics{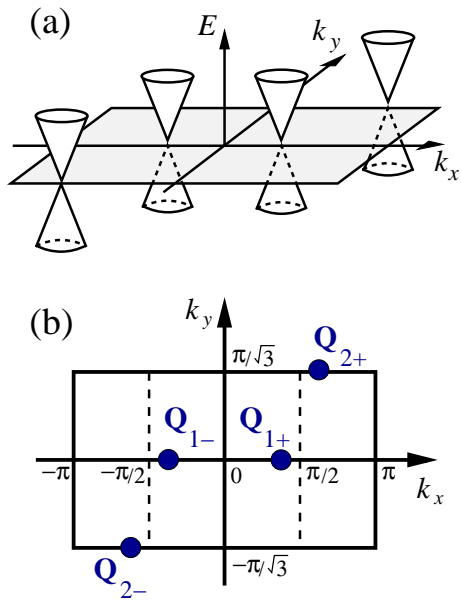}}}  
  \end{center}  
  \caption{(a) Schematic energy dispersion with $t'/t<\sqrt{2}$  
  near the four Dirac points which compose the Fermi ``surface'' for  
  the half-filled fermions in the flux-smeared mean-field state.   
  (b) Locations of the Dirac points in the rectangular Brillouin zone 
  corresponding to our unit cell choice in Fig.\ \ref{UnitCell}.   
  The pairs of nodes ${\bf Q}_{1+}$, ${\bf Q}_{1-}$ and ${\bf 
  Q}_{2+}$, ${\bf Q}_{2-}$ coalesce  
  when $t'/t = \sqrt{2}$ and become gapped for $t'/t>\sqrt{2}$. }  
  \label{DiracPts}  
\end{figure}

Working out the fermionized-vortex band structure for ${\mathcal 
H}_{\rm MF}$ exposes the important low-energy degrees of freedom in the 
mean-field theory.  Doing so will enable us to derive a continuum 
mean-field Hamiltonian, which will serve as the foundation on 
which we construct the full interacting low-energy theory by restoring 
fluctuations about the flux-smeared mean-field state.   
To this end, we diagonalize ${\mathcal H}_{\rm MF}$ in momentum space  
assuming the four-site unit cell shown in Fig.\ \ref{UnitCell},  
choosing a gauge with  
$a^0_{\bf x x'} = \pi/4$ directed along the arrows in the figure. 
Throughout we take as our origin a triangular lattice site denoted by 
the filled circle in Fig.\ \ref{UnitCell}.   
The band structure consists of four bands, two with positive energy 
and two with negative energy.   
Explicitly, the band energies at wave vector $(k_x,k_y)$ in the Brillouin 
zone of Fig.~\ref{DiracPts} are given by 
\begin{eqnarray*} 
E^2 = t^{\prime 2} + 2t^2 \pm 2t\sqrt{t^2 \sin^2 k_x + 
t^{\prime 2} [1+\cos k_x \cos(\sqrt{3}k_y)]} ~. 
\end{eqnarray*} 
The spectrum has Dirac nodes at zero energy for $t'/t<\sqrt{2}$ and  
is gapped for $t'/t > \sqrt{2}$; we discuss these two cases separately 
below.  Our main focus will be on the former gapless regime,  
where the algebraic vortex liquid arises.   
In the latter case with gapped vortices, which corresponds to 
the ``square-lattice'' Neel phase shown in Fig.~\ref{PhaseDiagram}, 
we will briefly discuss how within our flux-smeared mean-field treatment 
we can account for the anomalous ``roton'' minima in the  
excitation spectra observed by series expansion  
studies.\cite{Zheng99,ZhengSeriesExpansion}

\subsubsection{$t'/t > \sqrt{2}$: Gapped vortices. \\ ``Rotons'' in the  
frustrated square-lattice} 
\label{SquareLattice}

\begin{figure}  
  \begin{center}  
    {\resizebox{6cm}{!}{\includegraphics{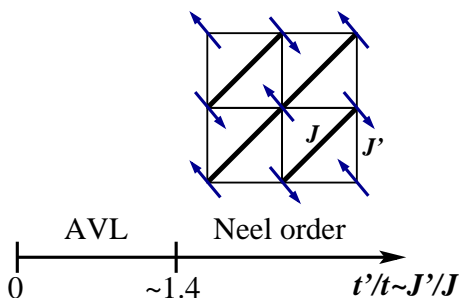}}}  
  \end{center}  
  \caption{Phase diagram for the dual fermionized-vortex 
  Hamiltonian in the flux-smeared mean-field state.  In the range of 
  anisotropy $t'/t > \sqrt{2}$, it is useful to view the system as 
  a square lattice antiferromagnet with nearest-neighbor exchange $J'$ and 
  frustrating coupling $J$ along one of the diagonal links as shown above.}  
  \label{PhaseDiagram}  
\end{figure}

For $t'/t> \sqrt{2}$, the half-filled fermionic vortices form a band 
insulator with the minimum band gap occurring at wave vectors  
${\bf Q}_1 = (0,0)$ and ${\bf Q}_2 = (\pi,\pi/\sqrt{3})$ in the 
Brillouin zone of Fig.\ \ref{DiracPts}. 
It is useful in this range of anisotropy to view the triangular system 
as a square lattice antiferromagnet with nearest-neighbor  
exchange $J'$ and frustrating antiferromagnetic exchange $J$ along one  
diagonal direction as shown in Fig.\ \ref{PhaseDiagram}.   
The vortex insulator realized here corresponds to the 
square-lattice Neel state, which for 
sufficiently small $J$ is the anticipated ground state.  Since we are 
considering an easy-plane model, the spins order in the $(S^x,S^y)$ 
plane.  The gapless Goldstone spin-wave at zero momentum is realized in 
the dual theory as a propagating ``photon'' mode in the 
electromagnetic gauge fields.  (The Goldstone mode at the ordering 
wave vector present in a Heisenberg system acquires a gap in the 
easy-plane limit.)  The Neel phase survives down to $t'/t = \sqrt{2}$, 
at which point the spectrum becomes gapless at ${\bf Q}_{1,2}$ signaling 
the destruction of the Neel order.   
Series expansion studies\cite{Zheng99} for the spatially  
anisotropic Heisenberg system find that  
the Neel phase survives in a similar range of anisotropy,  
$J'/J \gtrsim 1.4$.

Interestingly, excitation spectra for the Neel state  
calculated in series expansion studies of the Heisenberg system  
show significant deviations from spin-wave  
theory.\cite{Zheng99,ZhengSeriesExpansion}   
In terms of the standard square-lattice Brillouin zone notation,  
linear spin wave theory  
predicts identical excitation energies at momenta $(\pi/2,\pi/2)$  
and $(\pi,0)$ irrespective of the frustrating coupling $J$.  Series  
expansions, on the other hand, obtain large  
energy differences between these momenta due to a ``roton'' minimum in 
the spectrum at $(\pi,0)$ which deepens as $J$ increases  
(see Fig.\ 3 in Ref.\ \onlinecite{ZhengSeriesExpansion}). 
When $J'/J = 1.7$, the excitation energy at $(\pi,0)$ is roughly 
27\% lower than that at $(\pi/2,\pi/2)$.\cite{Zheng99,ZhengSeriesExpansion}   
 
The anomalous minimum can be accounted for within our  
flux-smeared mean-field treatment as a low-energy vortex-antivortex 
excitation, thus substantiating the roton interpretation. 
Before proceeding, we want to note that  
Refs.~\onlinecite{Zheng99, ZhengSeriesExpansion} consider $S^z = 1$  
excitations since the projection of the 
total spin onto the Neel vector direction (assumed to be along $S^z$)  
is conserved in the ordered phase of  
the Heisenberg system.  There is no such spin quantum number in the 
easy-plane case, and we can characterize the excitations only by their 
momenta.  If necessary, in the Heisenberg case there are always  
low-energy magnons 
near zero momentum and $(\pi,\pi)$ which can be added  
to the vortex-antivortex  
excitations discussed below to get the required spin quantum number. 
 
\begin{figure}  
  \begin{center}  
    {\resizebox{7cm}{!}{\includegraphics{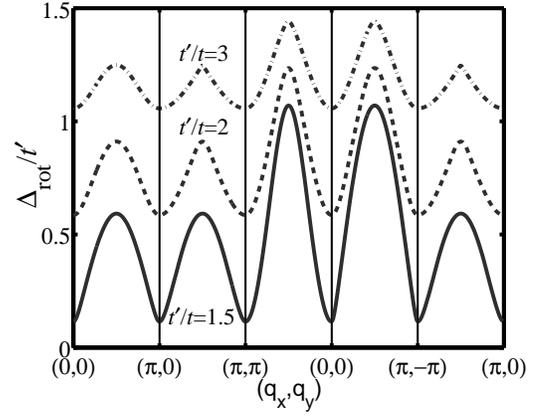}}}  
  \end{center}  
  \caption{``Roton'' excitation energy $\Delta_{\rm rot}({\bf q})$ in the Neel 
  phase along various cuts in momentum space.  To facilitate 
  comparison with Fig.\ 3 from Ref.\ 
  \onlinecite{ZhengSeriesExpansion}, we use the same square 
  lattice wave vector notation.  As frustration increases, the roton excitation 
  energy at $(\pi,0)$ decreases, which is consistent with series 
  expansion studies of the Heisenberg  
  system,\cite{Zheng99,ZhengSeriesExpansion}  
  and eventually become \emph{gapless} as the Neel 
  order is destroyed.  }  
  \label{Roton}  
\end{figure}

To work more formally, consider the dynamical correlation of the $S^z$ 
operator at momentum ${\bf q}$ in the flux-smeared mean field theory. 
$S^z_{\bf q}$  
obtains contributions from vortex 
currents whose circulation induces flux in the dual gauge field. 
The precise form of these vortex currents will be unimportant here,  
but can be obtained by 
constructing perturbations to the hopping Hamiltonian which give rise 
to static gauge flux modulated at wave vector ${\bf q}$.   
Generically, these contributions can be expressed as 
\begin{equation} 
  S^z_{\bf q} \sim \sum_{a,b = 1}^4\sum_{\bf k}\gamma_{a,b}({\bf k},{\bf q})  
  d_a^\dagger({\bf k})d_b({\bf k}-{\bf q}). 
  \label{Szq} 
\end{equation} 
Here $a,b$ are band indices, ${\bf k}$ is summed over the Brillouin 
zone in Fig.\ \ref{DiracPts}(b), $d_a^\dagger({\bf k})$ adds a  
fermion with momentum ${\bf k}$ in band $a$, and $\gamma_{a,b}$ are 
generally nonvanishing complex factors.  Consequently, the spin  
structure factor has contributions not only from spin-waves, but also 
from vortex-antivortex ``roton'' excitations.   
In the ground state the two lower bands are filled, while the upper 
bands are empty.  The 
excitation energy $\Delta_{\rm rot}({\bf q})$ for a roton with momentum  
${\bf q}$ is thus simply given by the minimum energy required to 
promote a fermion with arbitrary momentum ${\bf k}$ from an occupied  
band to a state with momentum ${\bf k-q}$ in an unoccupied band: 
\begin{equation} 
  \Delta_{\rm rot}({\bf q}) = \min_{\bf k} \{E_{\rm empty}({\bf k-q})  
  - E_{\rm filled}({\bf k}) \} ~. 
\end{equation}  
This is straightforward to compute from the vortex band structure.   
The result for $\Delta_{\rm rot}({\bf q})$ along several cuts in  
momentum space is shown in Fig.\ \ref{Roton} for three values of $t'/t$. 
To facilitate comparison with the series expansion results,  
we use square-lattice notation in the figure and display the same  
cuts as in  
Fig.~3 of Ref.~\onlinecite{ZhengSeriesExpansion}. 
Note that unlike Ref.~\onlinecite{ZhengSeriesExpansion} 
which shows the lower edge for all excitations including spin waves, 
our figure shows only the vortex-antivortex excitations. 
We also point out that the vortex-antivortex excitation energies satisfy 
$\Delta_{\rm rot}({\bf q}) = \Delta_{\rm rot}({\bf q}+(\pi,0))  
= \Delta_{\rm rot}({\bf q}+(0,\pi)) $, 
though this does not hold for general excitations. 
 
The most notable feature to observe in Fig.\ \ref{Roton} is that 
the roton excitation energy at $(\pi,0)$ decreases as the frustration 
increases, consistent with series expansions, and becomes  
\emph{gapless} as the Neel order gets destroyed.  The roton 
excitations at $(0,0)$ and $(\pi,\pi)$ follow the same trend, though 
these low-energy rotons would be difficult to observe in a Heisenberg 
system due to the gapless spin-waves at these momenta.  However, since 
spin-waves at $(\pi,\pi)$ are gapped in the easy-plane limit, 
significant deviations from spin-wave theory due to the roton at this  
wave vector are expected.  Series expansions for an easy-plane system 
to search for this anomaly would be interesting, and could serve as a 
test for our explanation of the Heisenberg spectra.

\subsubsection{$t'/t <\sqrt{2}$: Critical vortices } 
\label{BandStructure} 
 
For $\tilde t \equiv t'/t<\sqrt{2}$ one finds that  
the Fermi ``surface'' for the half-filled fermionic vortices  
consists of four gapless, linearly 
dispersing Dirac points shown schematically in Fig.\ \ref{DiracPts}(a).   
With our gauge choice these Dirac points occur at generally  
incommensurate wave vectors ${\bf Q}_{1\pm}$ and ${\bf Q}_{2\pm}$  
which can be written 
\begin{eqnarray} 
  {\bf Q}_{1\pm} &=& \pm(\pi/2-\tilde Q,0) ~, 
  \\ 
  {\bf Q}_{2\pm} &=& \pm(\pi/2+\tilde Q,\pi/\sqrt{3}) ~, 
\end{eqnarray} 
with 
\begin{equation} 
  \tilde Q \equiv \frac{\pi}{2} - 
  \cos^{-1}\bigg{(}\frac{\tilde t^2}{2}\bigg{)}. 
  \label{Qtilde} 
\end{equation} 
Fig.\ \ref{DiracPts}(b) shows the positions of these wave vectors in the  
rectangular Brillouin zone corresponding to our unit cell choice. 
The Dirac points have the familiar relativistic dispersion $E \approx 
\pm \sqrt{v_x^2 q_x^2 + v_y^2 q_y^2}$, where ${\bf q}$ is measured 
relative to the nodal wave vectors ${\bf Q}_{Ll}$ ($L = 1,2; l = +,-$).   
The velocities $v_{x,y}$ are in general 
anisotropic due to the anisotropy in the hopping amplitudes $t_{\bf x 
x'}$ and are given by 
\begin{eqnarray} 
  v_x &=& t\bigg{(}1-\frac{\tilde t^2}{2}\bigg{)}^{1/2} ~, 
  \\ 
  v_y &=& v_x \bigg{(}\frac{3\tilde t^4}{4-\tilde t^4}\bigg{)}^{1/2} ~. 
\end{eqnarray} 
Note some limiting cases:  
in the isotropic case $\tilde t = 1$ the velocities are equal; 
in the 1D limit $\tilde t \rightarrow 0$ the spectrum becomes  
dispersionless in the $y$-direction; and finally, as $\tilde t 
\rightarrow \sqrt{2}$ and  
we approach the square lattice Neel state, the Dirac cones merge in pairs and 
flatten in the $x$-direction, with $v_x\rightarrow 0$.

For the purpose of exploring the low-energy physics of the theory, it 
suffices to focus only on low-energy excitations in the vicinity of 
the Dirac nodes.  This can be achieved by expanding the fermion operators 
around the wave vectors ${\bf Q}_{Ll}$ as follows, 
\begin{equation} 
  d_{{\bf x},\ell} \approx \sum_{Ll \alpha} e^{i {\bf Q}_{Ll}\cdot 
  {\bf x}} \Phi^{Ll\alpha}_\ell \psi_{Ll\alpha}, 
  \label{psi} 
\end{equation} 
where ${\bf x}$ denotes sites of the honeycomb as before and  
$\ell=1,\ldots,4$ labels the corresponding site index in the unit cell 
pictured in Fig.\ \ref{UnitCell}.  
On the right side of Eq.\ (\ref{psi}), $\psi_{Ll}$ are two-component  
spinors assumed to  
vary slowly on the lattice scale, with spinor components indexed by  
the label $\alpha = \uparrow,\downarrow$. 
Up to an overall uniform normalization factor,  
the ``eigenvectors'' $\Phi^{Ll\alpha}$ can be written  
\begin{eqnarray} 
  \Phi^{1+\uparrow} =  
  \begin{bmatrix} 
    -\tilde t \\ s \\ 0 \\ 0 
  \end{bmatrix} 
  &;& \Phi^{1+\downarrow} =  
  \begin{bmatrix} 
    0 \\ 0 \\ -\tilde t \\ s 
  \end{bmatrix} 
  \nonumber \\ 
  \Phi^{1-\uparrow} =  
  \begin{bmatrix} 
    0 \\ 0 \\ s \\ -\tilde t 
  \end{bmatrix} 
  &;& \Phi^{1-\downarrow} =  
  \begin{bmatrix} 
    -s \\ \tilde t \\ 0 \\ 0 
  \end{bmatrix} 
  \nonumber \\ 
  \Phi^{2+\uparrow} = e^{-i\frac{\pi}{12}}  
  \begin{bmatrix} 
    0 \\ 0 \\ -i\tilde t \\ s 
  \end{bmatrix} 
  &;& \Phi^{2+\downarrow} = e^{i\frac{\pi}{12}} 
  \begin{bmatrix} 
    -i\tilde t \\ -s \\ 0 \\ 0  
  \end{bmatrix} 
  \nonumber \\ 
  \Phi^{2-\uparrow} = e^{-i\frac{\pi}{12}} 
  \begin{bmatrix} 
    s \\ -i \tilde t \\ 0 \\ 0 
  \end{bmatrix} 
  &;& \Phi^{2-\downarrow} = e^{i\frac{\pi}{12}}, 
  \begin{bmatrix} 
    0 \\ 0 \\ s \\ i\tilde t 
  \end{bmatrix} 
\end{eqnarray} 
where $s \equiv 2 \sin(\tilde Q/2)$. 
   
Using the expansion for the fermion operators in Eq.\ (\ref{psi}),  
we obtain the following low-energy continuum description for the mean-field 
Hamiltonian,  
\begin{equation} 
  \mathcal{H}_{MF} \sim \int d{\bf x}\; \psi^\dagger_{Ll}(-i v_x \partial_x 
  \sigma^x -i v_y \partial_y \sigma^y) \psi_{Ll}, 
  \label{Htcontinuum} 
\end{equation} 
where the flavor indices $Ll$ are implicitly summed and  
$\sigma^{x,y}$ are Pauli matrices that contract with the spinor  
indices [{\it i.e.},  
$(\sigma^x\psi)_{Ll\alpha} \equiv \sigma^x_{\alpha\beta}\psi_{Ll\beta}$]. 
Proceeding to the imaginary-time path integral 
formulation, the Euclidean Lagrangian 
density obtained from Eq.\ (\ref{Htcontinuum}) can be written 
\begin{eqnarray} 
  \mathcal{L}_{MF} &\sim& \overline \psi_{Ll} \gamma^\mu \partial_\mu 
  \psi_{Ll}, 
  \\ 
  \overline \psi_{Ll} &\equiv& \psi^\dagger_{Ll}\gamma^0, 
\end{eqnarray} 
where the space-time index $\mu = 0,1,2$ is defined so that 
$\partial_{0,1,2} \equiv \partial_{\tau,x,y}$ and  
we have rescaled the spatial coordinates to absorb the 
anisotropic velocities $v_{x,y}$.  The Dirac matrices $\gamma^\mu$ are 
given by $\gamma^0 = \sigma^z$, $\gamma^1 = \sigma^y$, $\gamma^2 = 
-\sigma^x$ and satisfy the usual algebra $\{\gamma^\mu,\gamma^\nu\} = 
2 \delta^{\mu\nu}$.  We will also frequently use Pauli matrices 
$\mu^k$ and $\tau^k$ which contract with the flavor indices $L$ and 
$l$, {\it i.e.}, 
\begin{eqnarray} 
  (\mu^k\psi)_{Ll} &\equiv& \mu^k_{LM}\psi_{Ml}  
  \nonumber \\ 
  (\tau^k\psi)_{Ll} &\equiv& \tau^k_{lm}\psi_{Lm}. 
  \label{FlavorMatrices} 
\end{eqnarray} 
 
Upon resurrecting the vortex interactions and gauge field 
fluctuations about the mean-field state, we obtain the 
desired low-energy theory, 
\begin{eqnarray}  
  {\mathcal L} &=&  
  \barpsi_{Ll} \gamma^\mu (\partial_\mu - i a_\mu - i A_\mu )\psi_{Ll}  
  \nonumber \\ 
  &+& \frac{1}{2 e^2} (\epsilon_{\mu\nu\lambda}\partial_\nu a_\lambda)^2  
  + \frac{i}{4\pi} \epsilon_{\mu\nu\lambda}A_\mu \partial_\nu A_\lambda  
  + {\mathcal L}_{\rm 4f}. 
  \label{L0}  
\end{eqnarray}  
Equation (\ref{L0}) describes four flavors of two-component Dirac fermions 
$\psi_{Ll}$ (corresponding to the four Dirac points)  
minimally coupled to a non-compact U(1) gauge field $a_\mu$ and a 
Chern-Simons field $A_\mu$.   
The gauge field $a_\mu$ mediates the logarithmic vortex repulsion,  
while the Chern-Simons 
terms above enforce the flux attachment to the fermions, thereby 
restoring the original bosonic vortex exchange statistics.  The 
form of the Maxwell term above is only schematic and ignores the 
underlying lattice anisotropy. 
Finally, $\mathcal{L}_{\rm 4f}$ 
represents symmetry-allowed four-fermion terms arising from  
short-range parts of the vortex interactions in the microscopic model.   
We furnish 
an explicit form of $\mathcal{L}_{\rm 4f}$ in Sec.\ \ref{NearbyPhases}.   
 
\begin{table*} 
\caption{\label{tab:psi} Transformation properties of the continuum 
fermion fields $\psi$.  Symmetries $T_1$ and $T_2$ correspond to 
translations by ${\bf \delta r} = {\bf \hat x}$ and ${\bf \delta r} = 
-1/2 {\bf \hat x} + \sqrt{3}/2 {\bf \hat y}$, respectively.   
Moreover,  
${\mathcal T}_{\rm ferm}$ corresponds to the naive time reversal for  
fermions on the honeycomb rather than the physical spin time reversal.  } 
\begin{ruledtabular}  
\begin{tabular}{c | c | c | c | c | c | c | c }  
  & $T_1$ & $T_2$ & $\tilde {\cal R}_x$ & $R_\pi$  
  & ${\mathcal C}$ & ${\cal T}_{\rm ferm}$ & $R_{\pi/3}$ (isotropic limit) 
  \\ \hline  
  $\psi \to$   
  & $i\tau^z e^{-i \tilde Q \mu^z \tau^z} \psi$  
  & $-i\mu^x \tau^y e^{i(\frac{\tilde Q}{2}-\frac{\pi}{4})\mu^z \tau^z} \psi$  
  & $e^{i\frac{\pi}{4}(\mu^z-1)} \psi$  
  & $\tau^x \sigma^z  \psi$ 
  & $\mu^x \tau^z \sigma^x [\psi^\dagger]^t$   
  & $\mu^y \sigma^y \psi$  
  & $e^{-i\frac{\pi}{6}\sigma^z} \mu^x e^{i\frac{\pi}{4} \mu^z} 
    e^{i\frac{\pi}{4} \mu^x \tau^x} \psi$  
\end{tabular}  
\end{ruledtabular}  
\end{table*}  
 
Table \ref{tab:psi} displays the transformation properties of the continuum  
fermion fields under the microscopic symmetries 
(with ${\mathcal T}_{\rm ferm}$ rather than ${\mathcal T}$ due to 
subtleties mentioned above). 
With out gauge choice for  $a^0_{\bf x x'}$ in Fig.~\ref{UnitCell},  
the two translations  
given in Table \ref{tab:psi} are realized as follows.  The first, 
$T_1$, corresponds to a simple translation of fields by  
$\delta {\bf r} = {\bf \hat x}$.  The second, $T_2$, 
corresponds to translation by ${\bf \delta r} = -1/2 {\bf \hat x} + 
\sqrt{3}/2 {\bf \hat y}$ and must be accompanied by a gauge transformation.   
Specifically, $T_2$ is realized by first implementing the required 
gauge transformation by sending  
\begin{eqnarray} 
  d_{{\bf x},1/4} &\to& i e^{i\pi(n_x+n_y)} d_{{\bf x},1/4} 
  \\ 
  d_{{\bf x},2/3} &\to& e^{i\pi(n_x+n_y)} d_{{\bf x},2/3}, 
\end{eqnarray} 
where $n_{x,y}$ are integers labeling the unit cell to which site 
${\bf x}$ belongs, and then translating as follows, 
\begin{eqnarray} 
  d_{{\bf x},1/2} &\to& d_{{\bf x+\delta r},2/1} 
  \\ 
  d_{{\bf x},3/4} &\to& d_{{\bf x+\delta r},4/3}. 
\end{eqnarray} 
Particle-hole symmetry and fermionic time reversal similarly require gauge 
transformations, as do rotations in the isotropic limit $J = J'$.   
 
Translations, reflections, and particle-hole symmetry prohibit all 
possible fermion bilinears from appearing  
in Eq.\ (\ref{L0}) except  
$i\barpsi \mu^z \tau^z \sigma^y \psi  
= \psi^\dagger \mu^z \tau^z \sigma^x \psi$ and 
$\barpsi \psi = \psi^\dagger \sigma^z \psi$. 
The first of these bilinears is a perturbation to the Hamiltonian which  
modifies the ratio of vortex hopping 
amplitudes $t'/t$.  This has the trivial effect of shifting the  
$x$-components of the nodal wave vectors ${\bf Q}_{Ll}$,  
which are not protected in an anisotropic system.  In the isotropic 
limit, $i\barpsi \mu^z \tau^z \sigma^y \psi$ is eliminated by rotation 
symmetry.  Deducing the fate of $\barpsi \psi$, which respects all 
symmetries except ${\mathcal T}_{\rm ferm}$, requires more care and 
will be discussed in Sec.\ \ref{AVLstability}. 
We will argue that adding this term to the action drives  
the system into the Kalmeyer-Laughlin chiral spin-liquid, which breaks  
physical time-reversal symmetry.  Thus $\barpsi \psi$ should be  
excluded if we are to describe a time-reversal invariant state.

At this point it is worth emphasizing that upon considering the simplest 
flux-smeared state, we have already arrived at a mean-field description  
of the ``critical'' algebraic vortex liquid which is the main focus  
of this paper.   
The ``critical'' nature of this state follows from the  
gaplessness of the fermionic 
vortices, which in turn allows for gapless spin excitations as we will 
discuss below. 
Many properties of the AVL, such as the momentum-space locations of 
the low-energy spin excitations, can in fact be deduced 
from the mean-field theory.  By studying the effective 
Lagrangian Eq.\ (\ref{L0}), we will attempt to go beyond such a 
mean-field analysis.  In particular, in the following section  
we will address the  
stability of the AVL when fluctuations about the flux-smeared state are 
incorporated and make quantitative predictions for various spin 
correlations in the AVL.  Moreover, with this effective theory in hand 
we can also explore the phase diagram in the vicinity of the AVL.   
The algebraic vortex liquid has a number of 
interesting proximate phases, some of which we explore in  
Sec.\ \ref{NearbyPhases}.  
 
\vskip .4cm 
To summarize, the mean-field phase diagram along the spatial 
anisotropy axis $t'/t$ is shown in Fig.\ \ref{PhaseDiagram}.  We will focus 
on $t'/t<\sqrt{2}$ for the remainder of the paper.

\section{Algebraic Vortex Liquid} 
\label{AVL}

\subsection{Effective theory of the AVL---QED3} 
\label{AVLstability} 
 
We begin our detailed characterization of the algebraic vortex liquid  
by examining the continuum theory describing fluctuations about the 
critical flux-smeared mean-field state. 
The full interacting theory is 
described by the effective Lagrangian Eq.\ (\ref{L0}).  To ascertain 
response properties of the spin system, we add an external probing 
field $A^{\rm ext}$ which couples to the  
three-current of the hard-core bosons.  In the dual vortex formulation 
prior to fermionization, this three-current is given by  
$\delta j = (\nabla \times 
a)/2\pi$.   Upon fermionization, we introduced an additional 
Chern-Simons field, and attached flux $(\nabla \times A)=2\pi$ to the vortices. 
As discussed at the beginning of Section \ III, 
it is convenient to view the Chern-Simons field 
as being ``part" of the original gauge field, that is  
$a \rightarrow \tilde{a} = a + A$. 
The idea is that it is the physical hard-core boson current which is becoming 
correlated with the motion of the vortices, forming a vortex-flux composite 
which behaves as a fermion.  Based on this picture, it is reasonable 
to assume that the Chern-Simons gauge flux carries the $U(1)$ charge of 
the hard-core bosons, and to couple in the external field via 
\begin{equation} 
  {\cal L}_{\rm ext} = -\frac{i}{2\pi}A^{\rm ext}\cdot[(\nabla\times 
    a)+(1-\kappa)(\nabla\times A)], 
  \label{deltaL} 
\end{equation} 
with $\kappa =0$.   
Traditional application of Chern-Simons theory 
would take instead $\kappa =1$, which corresponds to the assumption that 
the Chern-Simons flux is ``fictitious" rather than physical and  
hence carries no quantum numbers.  As we shall see below, the  
choice $\kappa=0$ is preferable, 
being essentially equivalent to replacing physical time reversal 
invariance by ${\mathcal T}_{\rm ferm}$.  But for now we keep  
$\kappa$ as an arbitrary parameter. 
 
Since the Dirac fermions in the low-energy continuum theory 
given in Eq.\ (\ref{L0}) couple only to $\tilde a_\mu = 
a_\mu + A_\mu$, it is instructive to rewrite the Lagrangian in terms of 
this sum field.  The full Lagrangian can then be cast in the following appealing form:  
\begin{equation} 
  {\cal L} = {\cal L}_{\rm QED3} + {\cal L}_{\rm CS}  
  + {\cal L}_{\rm int} + {\cal L}_{\rm ext}, 
\end{equation} 
with 
\begin{eqnarray} 
  {\mathcal L}_{\rm QED3} &=& \barpsi_{Ll} 
                  \gamma^\mu (\partial_\mu - i \tilde{a}_\mu )\psi_{Ll} 
  + \frac{1}{2 e^2} (\nabla \times \tilde a)^2 
   \nonumber \\ 
  &&+   {\cal L}_{4f}  
 ~ , 
 \label{QED3} 
\end{eqnarray}  
and 
\begin{equation} 
  {\cal L}_{\rm CS} = \frac{i}{4\pi} A\cdot(\nabla \times A) . 
  \end{equation} 
Here, ${\cal L}_{\rm QED3}$ describes 
non-compact quantum electrodynamics in $2+1$ dimensions (QED3) 
with $N=4$ flavors, which is coupled to the  
Chern-Simons Lagrangian by an interaction,  
\begin{equation} 
  {\cal L}_{\rm int} =  
  -\frac{1}{e^2} (\nabla \times \tilde a) \cdot (\nabla \times A) . 
\end{equation} 
The external probing field takes the form 
\begin{equation} 
  {\cal L}_{\rm ext} =   
  -\frac{i}{2\pi}A^{\rm ext}\cdot(\nabla \times \tilde a)  
  + \kappa\frac{i}{2\pi}A^{\rm ext}\cdot(\nabla \times A). 
\end{equation} 
Notice that with the choice $\kappa=0$, the external source field only 
couples to $\tilde{a}$. 
  
Before discussing the effects of the interaction term, we briefly review 
the behavior of QED3, which has been widely studied in a variety of  
contexts.\cite{Appelquist, Rantner, Vafek, Franz, Herbut, Hermele, 
  spin1, QED3sim1, QED3sim2}    
The fixed point with $e^2 = 0$ in which  
gauge fluctuations are entirely suppressed and the fermions are 
essentially free is unstable, so that QED3 is inherently a  
strongly interacting  
field theory.  Consequently, to make progress analytically one must  
modify the theory in a manner which provides a controlled limit.   
An often used approach is the large-$N$ limit, where one generalizes  
to a large number $N$ of fermion flavors.  Starting from the 
infinite-$N$ limit, one can then perform a controlled analysis 
by perturbing in powers of $1/N$.   
This approach can be cast in the form of a renormalization group 
treatment, and an important feature is that the gauge field 
$\tilde{a}$ scales like an inverse length and due to gauge  
invariance does not pick up an anomalous dimension. 
The scaling dimensions of the symmetry-allowed 
four-fermion interactions in Eq.\ (\ref{QED3}) 
do generally acquire an anomalous dimension, which can be computed 
perturbatively in inverse powers of $N$. 
For large enough $N$  
all four-fermion terms are irrelevant, and QED3 thus realizes a 
nontrivial stable critical phase.   
For $N<N_c$, with some unknown $N_c$, it is believed that 
four-fermion terms become relevant, leading to spontaneous fermion 
mass generation and the destruction of criticality (except with 
fine-tuning).  While $N_c$ can in principle be deduced in 
QED3 simulations, recent studies are inconclusive as to whether $N_c$ 
lies above or below the $N = 4$ case of interest in the present  
work.\cite{QED3sim1, QED3sim2}   
We will assume henceforth that $N_c < 4$.  Further numerical simulations and 
higher-order calculations in $1/N$ would be useful for justifying (or 
negating) this assumption.

We are now in position to consider the effects of the coupling 
${\cal L}_{\rm int}$ between QED3 and the Chern-Simons Lagrangian. 
Since ${\cal L}_{\rm CS}$ is Gaussian, it can be viewed as the fixed point 
of a renormalization group transformation in which $A$ is rescaled 
like an inverse length.  For large $e$, the effects of the interaction can 
then be studied perturbatively.  Since ${\cal L}_{\rm int}$ is quadratic 
in the gauge fields and involves two derivatives it has scaling 
dimension 4, and is formally irrelevant.  The fixed points described 
by QED3 and Chern-Simons theory evidently decouple at 
low energies.   The physics here is that due to the logarithmic vortex 
repulsion which strongly suppresses vortex density fluctuations,  
exchange statistics play only a minor role at criticality. 
In the next subsection we will employ QED3 to access the properties 
of the critical algebraic vortex liquid phase. 
 
But caution is necessary.  For ${\cal L}_{\rm QED3}$  
fermionic time reversal symmetry precludes the generation of a  
fermionic mass term $\barpsi \psi$, which 
respects all symmetries of the Lagrangian except ${\mathcal T}_{\rm ferm}$. 
The Chern-Simons Lagrangian ${\cal L}_{\rm CS}$, however, is not  
invariant under ${\mathcal T}_{\rm ferm}$.  As a result, 
once the two theories are coupled, despite the irrelevance of this 
coupling a small fermion mass term will presumably be generated, being  
no longer symmetry-protected.  Tracing back its origin, we  
see that the sign of the vortex mass will be determined by the  
sign of the flux that was attached upon vortex fermionization.   
As we will discuss in Sec.\ \ref{KLsl}, this mass 
term drives the system into the Kalmeyer-Laughlin chiral spin-liquid, 
which breaks \emph{physical} spin time-reversal symmetry.   
Thus, in order to correctly implement a renormalization group analysis 
that faithfully respects the physical time reversal symmetry of  
the original spin model, we must maintain masslessness of the fermions.   
We will proceed under the assumption that the physically correct procedure is 
to tune a small bare mass term to cancel the effects 
of the irrelevant coupling as it scales to zero---that is, to tune 
the fully renormalized mass term to zero.   
The resulting massless 
and critical QED3 gives us a description of the time reversal  
invariant algebraic vortex liquid. 
 
Subtleties associated with time reversal invariance  
are also apparent in the Hall conductivity 
of the original hard-core bosons, which we now briefly discuss. 
Once the Chern-Simons Lagrangian ${\cal L}_{\rm CS}$ has decoupled, one can  
readily perform the Gaussian integration over $A$, which gives, 
\begin{equation} 
  {\cal L}_{\rm ext} =  
  -\frac{i}{2\pi}A^{\rm ext}\cdot(\nabla \times \tilde a) 
  -\kappa^2\frac{i}{4\pi}A^{\rm ext}\cdot(\nabla \times A^{\rm ext}). 
\label{Lext2} 
\end{equation} 
This form shows that the conductivity tensor of the 
original hard-core bosons,  $\sigma_{\alpha\beta}$, is given by 
\begin{equation} 
  \sigma_{\alpha\beta} = \frac{\rho^{\rm fv}_{\alpha\beta}}{(2\pi)^2}  
+ \frac{\kappa^2}{2\pi} \epsilon_{\alpha\beta}, 
  \label{sigma} 
\end{equation} 
where $\rho^{\rm fv}$ is the resistivity tensor for the fermionic  
vortices described by QED3 and $\epsilon$ is the antisymmetric tensor.   
Since the Hall resistivity for the fermions vanishes in QED3 
due to ${\mathcal T}_{\rm ferm}$ symmetry, the Hall conductivity 
for the bosons in the critical AVL phase is given by  
$\sigma_{xy} = \kappa^2/2\pi$. 
Recall that the parameter $\kappa$ gives a measure of how much 
bosonic charge is ascribed to the statistical flux attached to the vortices.

Generally,  
the Hall conductivity is \emph{not} a low-energy  
property of a physical system, and as such can be non-universal 
even at a critical point or in a critical phase.  However, since the  
original spin model 
is time reversal invariant, the Hall conductivity must vanish, 
at least in the absence of any spontaneous symmetry breaking. 
Notice that the required vanishing of the Hall conductivity 
follows provided we take the parameter $\kappa=0$. 
As discussed above and in Section \ III, 
the choice $\kappa=0$ corresponds to assuming that the 
Chern-Simons flux attached to the vortices carries a non-vanishing 
boson charge. 
On the other hand, if the statistical flux is presumed to carry no charge, 
one has $\kappa=1$ and a non-vanishing Hall conductivity, 
with a sign set by the sign of the attached statistical flux. 
The former choice, $\kappa=0$, gives us a way to 
access a time reversal invariant state with zero Hall conductivity 
independent of the sign of attached flux. 
Physically, with the fermions coupling to the sum $\tilde{a} = a+A$,  
the motion of the vortices leads to a ``screening" of the 
attached flux $\nabla \times A$, by the fluctuating flux $\nabla \times a$. 
At long wavelengths the total flux 
surrounding each vortex, $\nabla \times \tilde{a}$, which 
is proportional to the full boson current when $\kappa=0$, 
vanishes.  The vortices are charge neutral and the Hall effect vanishes. 
 
It would clearly be desirable to have a method for fermionization 
involving flux attachment in a more democratic fashion 
which treats $+2\pi$ and $-2\pi$ in an exactly equivalent manner. 
But in the absence of such an approach, we must content ourselves 
with using ${\cal L}_{\rm QED3}$ together with the 
assumption of  
$\kappa=0$ to describe the properties of  
the time-reversal invariant AVL phase.

\subsection{Symmetries of the AVL} 
\label{AVLsymmetries} 
 
\begin{figure}  
  \begin{center}  
    {\resizebox{6cm}{!}{\includegraphics{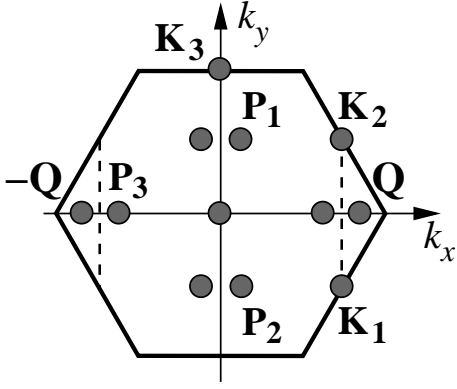}}}  
  \end{center}  
  \caption{Momenta carried by the fermion bilinears in Table 
  \ref{tab:EnhancedBilinears}, whose correlations are enhanced by 
  gauge fluctuations at the AVL fixed point.  }  
  \label{BilinearQs}  
\end{figure}

The critical QED3 theory proposed to describe the AVL  
respects all symmetries in Table 
\ref{tab:psi}, and also has a dual global U(1) symmetry 
under $\psi \rightarrow e^{i\alpha} \psi$ reflecting conservation of 
vorticity.  Due to the assumed irrelevance of four-fermion terms  
in the scaling limit, the theory also possesses   
an emergent global SU(4) flavor symmetry, being invariant under  
arbitrary SU(4) flavor rotations of the form $\psi \rightarrow U \psi$, 
with $U^\dagger = U^{-1}$.  The 16 conserved three-currents associated with the 
U(1) and SU(4) symmetries can be compactly written 
\begin{equation} 
  J^\nu_{\alpha\beta} = \barpsi \gamma^\nu \mu^\alpha \tau^\beta \psi 
\end{equation} 
and satisfy $\partial_\nu J^\nu_{\alpha\beta} = 0$. 
Here the indices $\alpha,\beta$ range from 0 to 3; $\mu^0$ and 
$\tau^0$ are identity matrices; and $\mu^j$ and $\tau^j$ are Pauli 
matrices defined as in Eq.\ (\ref{FlavorMatrices}).   
The U(1) conserved current is $J^\nu_{00}$, while 
the remaining 15 currents constitute the SU(4) conserved currents.  
The 48 fermion bilinears comprising $J^\nu_{\alpha\beta}$ 
are prohibited from acquiring an anomalous dimension.   
We will be primarily interested in the remaining 16 bilinears, whose  
correlations are  
enhanced by gauge fluctuations; their transformation properties are 
supplied in Table \ref{tab:EnhancedBilinears}.   
Figure \ref{BilinearQs} displays the set of momenta carried by these 
enhanced fermion bilinears, which correspond to the leading gapless 
vortex-antivortex excitations.  The wave vectors in the figure 
are explicitly given by $\pm {\bf Q}$, where  
\begin{equation} 
  {\bf Q} = (2\tilde Q +\pi,0) 
  \label{Qspiral} 
\end{equation} 
and $\tilde Q$ is defined in Eq.\ (\ref{Qtilde}), ${\bf K}_{1,2} = 
(\pi,\mp\pi/\sqrt{3})$, ${\bf K}_{3} = (0,2\pi/\sqrt{3})$, and 
$\pm{\bf P}_j = \pm({\bf Q}+{\bf K}_j)$.   
We will often refer to $\pm{\bf Q}$ as spiral ordering wave vectors, since 
in the isotropic limit these correspond to  
the $\sqrt{3} \times \sqrt{3}$ order.

\begin{table*} 
\caption{\label{tab:EnhancedBilinears} Transformation properties of 
the 16 bilinears whose correlations are enhanced by gauge fluctuations  
at the AVL fixed point (we do not show separately  
${\mathcal M}_{SS}^\dagger$ 
and ${\mathcal P}_{1,2,3}^\dagger$).  In the second column,  
${\bf Q}$ is the spiral ordering wave vector defined in Eq.~(\ref{Qspiral}),  
${\bf K}_j$ lie on the 
midpoints of the Brillouin zone edges, 
and ${\bf P}_j$ = ${\bf Q}+{\bf K}_j$.  Figure \ref{BilinearQs}
displays the set of momenta carried by the enhanced fermion bilinears.  }  
\begin{ruledtabular}  
\begin{tabular}{c | c | c | c | c | c | c}  
  & $T_{\delta{\bf r}}$ & $\tilde {\cal R}_x$ & $R_\pi$  
  & ${\mathcal C}$ & ${\cal T}_{\rm ferm}$ & $R_{\pi/3}$ (isotropic limit) 
  \\ \hline 
  $\barpsi \psi = {\mathcal M}_{KL} \to$ & ${\mathcal M}_{KL}$  
  & ${\mathcal M}_{KL}$ & ${\mathcal M}_{KL}$ & ${\mathcal M}_{KL}$ &  
  $-{\mathcal M}_{KL}$ & ${\mathcal M}_{KL}$  
  \\ \hline 
  $\barpsi \mu^z \tau^z \psi = {\mathcal M}_{\sqrt{3} \times \sqrt{3}} \to$  
  & ${\mathcal M}_{\sqrt{3} \times \sqrt{3}}$  
  & ${\mathcal M}_{\sqrt{3} \times \sqrt{3}}$  
  & $-{\mathcal M}_{\sqrt{3} \times \sqrt{3}}$  
  & $-{\mathcal M}_{\sqrt{3} \times \sqrt{3}}$  
  & ${\mathcal M}_{\sqrt{3} \times \sqrt{3}}$  
  & $-{\mathcal M}_{\sqrt{3} \times \sqrt{3}}$  
  \\ \hline 
  $\barpsi (\tau^x+i\mu^z\tau^y) \psi = {\mathcal M}_{SS} \to$  
  & $e^{i {\bf Q}\cdot {\bf \delta r}} {\mathcal M}_{SS}$ & ${\mathcal M}_{SS}$ 
  & ${\mathcal M}_{SS}^\dagger$ & $-{\mathcal M}_{SS}$  
  & $-{\mathcal M}_{SS}^\dagger$ & ${\mathcal M}_{SS}^\dagger$ 
  \\ \hline 
  $\barpsi \mu^x\tau^x \psi = {\mathcal K}_{1} \to$  
  & $e^{i {\bf K}_1\cdot {\bf \delta r}} {\mathcal K}_1$ & ${\mathcal K}_2$ & 
  ${\mathcal K}_1$ & $-{\mathcal K}_1$ & ${\mathcal K}_1$ & ${\mathcal K}_3$ 
  \\ \hline 
  $\barpsi \mu^y \tau^x \psi = {\mathcal K}_{2} \to$  
  & $e^{i {\bf K}_2\cdot {\bf \delta r}} {\mathcal K}_2$ & ${\mathcal K}_1$ & 
  ${\mathcal K}_2$ & $-{\mathcal K}_2$ & ${\mathcal K}_2$ & ${\mathcal K}_1$ 
  \\ \hline 
  $\barpsi \mu^z \psi = {\mathcal K}_{3} \to$  
  & $e^{i {\bf K}_3\cdot {\bf \delta r}} {\mathcal K}_3$ & ${\mathcal K}_3$ & 
  ${\mathcal K}_3$ & $-{\mathcal K}_3$ & ${\mathcal K}_3$ & ${\mathcal K}_2$ 
  \\ \hline 
  $\barpsi \mu^y\tau^y \psi = {\mathcal K}'_{1} \to$  
  & $e^{i {\bf K}_1\cdot {\bf \delta r}} {\mathcal K}'_1$  
  & $-{\mathcal K}'_2$ & $-{\mathcal K}'_1$ & ${\mathcal K}'_1$  
  & $-{\mathcal K}'_1$ & ${\mathcal K}'_3$ 
  \\ \hline 
  $\barpsi \mu^x\tau^y \psi = {\mathcal K}'_{2} \to$  
  & $e^{i {\bf K}_2\cdot {\bf \delta r}} {\mathcal K}'_2$  
  & $-{\mathcal K}'_1$ & $-{\mathcal K}'_2$ & ${\mathcal K}'_2$  
  & $-{\mathcal K}'_2$ & ${\mathcal K}'_1$ 
  \\ \hline 
  $\barpsi \tau^z \psi = {\mathcal K}'_{3} \to$  
  & $e^{i {\bf K}_3\cdot {\bf \delta r}} {\mathcal K}'_3$  
  & ${\mathcal K}'_3$ & $-{\mathcal K}'_3$ & ${\mathcal K}'_3$  
  & $-{\mathcal K}'_3$ & $-{\mathcal K}'_2$ 
  \\ \hline 
  $\barpsi(\mu^x+i\mu^y\tau^z)\psi = {\mathcal P}_1 \to $  
  & $e^{i {\bf P}_1\cdot {\bf \delta r}} {\mathcal P}_1$ & ${\mathcal P}_2$ & 
  ${\mathcal P}_1^\dagger$ & ${\mathcal P}_1$  
  & ${\mathcal P}_1^\dagger$ & ${\mathcal P}_3^\dagger$ 
  \\ \hline 
  $\barpsi(\mu^y-i\mu^x\tau^z)\psi = {\mathcal P}_2 \to $  
  & $e^{i {\bf P}_2\cdot {\bf \delta r}} {\mathcal P}_2$ & ${\mathcal P}_1$ & 
  ${\mathcal P}_2^\dagger$ & ${\mathcal P}_2$  
  & ${\mathcal P}_2^\dagger$ & ${\mathcal P}_1^\dagger$ 
  \\ \hline 
  $\barpsi(\mu^z\tau^x+i\tau^y)\psi = {\mathcal P}_3 \to $  
  & $e^{i {\bf P}_3\cdot {\bf \delta r}} {\mathcal P}_3$ & ${\mathcal P}_3$ & 
  ${\mathcal P}_3^\dagger$ & ${\mathcal P}_3$  
  & ${\mathcal P}_3^\dagger$ & ${\mathcal P}_2^\dagger$  
\end{tabular}  
\end{ruledtabular}  
\end{table*}  
 
\subsection{Dynamical spin correlations in the AVL} 
\label{SpinCorrelations} 
 
We turn now to the dynamical spin correlations in the AVL.  Due to the 
gaplessness of the fermionic vortices, the AVL admits universal 
power-law correlations in the spin structure factor.  To extract these  
spin correlations from our dual theory, we need to first identify the  
operators in QED3 which correspond to $S^z$ and $S^+$.  We  
discuss the correlations of $S^z$ and $S^+$ separately below.

\subsubsection{$S^{z}$ correlators} 
\label{Szz} 
 
From our microscopic identification in Eq.\ (\ref{Sz}), it is clear  
that near zero momentum $S^z$ appears in QED3 as the conserved  
dual gauge flux, $(\nabla \times a)/2\pi$.  Since $a$ is massless in the 
critical theory, $S^z$ exhibits power-law correlations at zero 
momentum with scaling dimension $2$.  At other wave vectors, $S^z$ receives 
contributions from fermion bilinears in QED3 which carry the same 
quantum numbers.   More precisely, due to subtleties with  
realizing physical time-reversal in QED3, we require that  
contributing fermion bilinears  
have the same quantum numbers as $\nabla \times a$. 
These bilinears arise microscopically from gapless vortex currents  
that induce gauge flux modulations at finite wave vector, which are 
analogous to the ``rotons'' discussed in Sec.\ \ref{SquareLattice}.   
 
From such an analysis, the continuum expression for $S^z$ takes the form 
\begin{equation} 
  S^z \sim \frac{\Delta\times a}{2\pi} +  
  [e^{i{\bf Q}\cdot {\bf r}}{\mathcal M}_{SS}+{\rm 
  H.c.}] +\cdots, 
  \label{SzContinuum} 
\end{equation} 
where ${\mathcal M}_{SS}$ and ${\mathcal M}_{SS}^\dagger$ are   
enhanced fermion bilinears from Table 
\ref{tab:EnhancedBilinears} that carry momenta $\pm{\bf Q}$.   
It can be readily verified  
using the table that the right-hand-side has the desired symmetry properties.   
The ellipsis 
in Eq.\ (\ref{SzContinuum}) represents terms arising from  
non-enhanced fermion bilinears  
({\it i.e.}, those which are part of the SU(4) conserved currents) and 
higher-order contributions.  For completeness, we note that these 
non-enhanced bilinears carry momenta ${\bf K}_{1,2,3}$ and $\pm {\bf 
  P}_{1,2,3}$ in Fig.\ \ref{BilinearQs}.   
Thus the $S^z$ correlations at momenta ${\bf K}_j$ and $\pm {\bf P}_j$ 
have scaling dimension $\Delta_{\rm nonenh} = 2$. 
 
Due to enhancement from gauge field fluctuations, the  
fermion bilinear ${\mathcal M}_{SS}$ in fact provides the dominant 
power law in the $S^z$ correlations.  Near momenta $\pm{\bf Q}$ 
the $S^{zz}$ spin structure factor scales as 
\begin{equation} 
  S^{zz}({\bf k} = \pm{\bf Q}+{\bf q},\omega) \sim  
  \frac{\Theta(\omega^2-{\bf q}^2)}{(\omega^2-{\bf q}^2)^{1-\eta_{\rm 
  enh}/2}}. 
\end{equation} 
The anomalous dimension $\eta_{\rm enh}$ is that of an enhanced 
fermion bilinear in QED3 and can be estimated from the  
leading $1/N$ result,\cite{Rantner}  
\begin{equation} 
  \eta_{\rm enh} \approx 3 - \frac{128}{3\pi^2 N}. 
\end{equation} 
Setting $N = 4$ yields $\eta_{\rm enh} \approx 1.92$, and a scaling 
dimension $\Delta_{\rm enh} = (1+\eta_{\rm enh})/2 \approx 1.46$.   
At all other wave vectors $S^{zz}$ exhibits subdominant power laws.   
 
The fact that the leading $S^z$ correlations occur at  
momenta $\pm{\bf Q}$ suggests proximity of 
``supersolids'' to the AVL.  Such states are characterized by 
concurrent $S^z$ and $S^+$ order, and are nontrivial in an easy-plane 
system where typically only in-plane spin order occurs.  As we will 
show in Sec.\ \ref{SSphases}, supersolid phases indeed emerge naturally out 
of the AVL.

\subsubsection{$S^{+}$ correlators} 
\label{S+-} 
 
Since the spin raising and lowering operators $S^\pm$ add spin $\pm1$,  
the corresponding operators in QED3 are ``monopole insertions'' which  
add $\pm2\pi$ gauge flux.  Our goal here will be to construct a 
continuum expression for $S^+$ in terms of these monopoles and from 
this extract the leading in-plane spin correlations, much as we 
did for $S^z$ above.  Monopole operators in QED3 were discussed 
in a very similar setting in Ref.\ \onlinecite{spin1}, and here we 
shall only highlight the main points.  We will assume that the 
added $\pm 2\pi$ flux is spread smoothly over a large area compared to 
the lattice unit cell, and treat the flux as a static 
background.  This flux alters the fermionic spectrum, and in 
particular gives rise to four zero-energy modes, one for each fermion 
flavor in the continuum.  The zero-mode wave functions can  
be obtained by first modifying  
the mean-field Hamiltonian density in Eq.\ (\ref{Htcontinuum}) as follows, 
\begin{equation} 
  {\mathcal H}_{MF,q} \sim -i\psi^\dagger_{Ll}[(\partial_x-ia^q_{x}) 
  \sigma^x +(\partial_y-ia^q_y) \sigma^y] \psi_{Ll}. 
\end{equation} 
Here $a^q_{x,y}$ is the vector potential giving rise to $2\pi q$ flux, 
with $q = \pm 1$ the monopole ``charge''.  Focusing only on the 
zero-modes, we then replace 
\begin{equation} 
  \psi_{Ll}({\bf x}) \rightarrow \phi_{Ll,q}({\bf x}) f_{Ll,q}, 
  \label{fAaq} 
\end{equation} 
where $\phi_{Ll,q}$ are the desired zero-mode wave functions and 
the operator $f_{Ll,q}$ annihilates the corresponding zero-mode.   
Choosing the Coulomb gauge for $a^q_{x,y}$ and assuming an azimuthally 
symmetric flux distribution centered around the origin, it is 
straightforward to show that the zero-mode wave functions are\cite{ZeroModes} 
\begin{eqnarray} 
  \phi_{Ll,+1} &\sim& \frac{1}{|{\bf x}|} \binom{1}{0} ~, 
  \\ 
  \phi_{Ll,-1} &\sim& \frac{1}{|{\bf x}|} \binom{0}{1}. 
\end{eqnarray} 
The transformation properties of the zero-mode operators $f_{Ll,q}$ 
can be deduced from Eq.\ (\ref{fAaq}) and the transformation 
properties of $\psi_{Ll}$ in Table \ref{tab:psi}; the results are shown in 
Table~\ref{tab:f}.   
 
Since the fermions are at half-filling, physical ({\it i.e.}, gauge-invariant) 
states must have two of the four zero-modes occupied.  Thus we need to  
consider six distinct monopole insertions.  It will be 
useful to define the following translation eigenoperators which add  
fermions to two of the four zero-modes: 
\begin{eqnarray} 
  F^\dagger_{0,q} &=& f_{1+,q}^\dagger f_{2+,q}^\dagger + 
  f_{1-,q}^\dagger f_{2-,q}^\dagger 
  \nonumber \\ 
  F^\dagger_{1,q} &=& f_{1+,q}^\dagger f_{1-,q}^\dagger + 
  f_{2+,q}^\dagger f_{2-,q}^\dagger 
  \nonumber \\ 
  F^\dagger_{2,q} &=& -i[f_{1+,q}^\dagger f_{1-,q}^\dagger - 
  f_{2+,q}^\dagger f_{2-,q}^\dagger] 
  \nonumber \\ 
  F^\dagger_{3,q} &=& f_{1-,q}^\dagger f_{2-,q}^\dagger - 
  f_{1+,q}^\dagger f_{2+,q}^\dagger 
  \nonumber \\ 
  F^\dagger_{R,q} &=& f_{1-,q}^\dagger f_{2+,q}^\dagger 
  \nonumber \\ 
  F^\dagger_{L,q} &=& f_{1+,q}^\dagger f_{2-,q}^\dagger. 
\label{Fop} 
\end{eqnarray} 
For convenience, the transformation properties of these operators are 
also displayed in Table~\ref{tab:f}.   
 
\begin{table*} 
\caption{\label{tab:f} Transformation properties of the zero-mode 
  operators $f_{Ll,q}$ in the charge $q = \pm 1$ monopole sectors. 
  The quoted transformations were obtained by employing the Coulomb gauge 
  for the added $\pm 2\pi$ gauge flux.   Also shown are the 
  transformation properties of the six translation eigenoperators  
  $F_{\alpha,q}^\dagger$ defined in Eq.~(\ref{Fop}) which add fermions  
  to two of the four zero-modes. } 
\begin{ruledtabular}  
\begin{tabular}{c | c | c | c | c | c | c | c }  
  & $T_1$ & $T_2$ & $\tilde {\cal R}_x$ & $R_\pi$  
  & ${\mathcal C}$ & ${\cal T}_{\rm ferm}$ & $R_{\pi/3}$ (isotropic limit) 
  \\ \hline  
  $f_q \to$   
  & $i\tau^z e^{-i \tilde Q \mu^z \tau^z} f_q$  
  & $-i\mu^x \tau^y e^{i(\frac{\tilde Q}{2} - \frac{\pi}{4})\mu^z \tau^z} f_q$  
  & $e^{i\frac{\pi}{4}(\mu^z-1)} f_q$  
  & $q \tau^x f_q$ 
  & $\mu^x \tau^z [f_{-q}^\dagger]^t$   
  & $-i q \mu^y f_{-q}$  
  & $e^{-i q \frac{\pi}{6}} \mu^x e^{i\frac{\pi}{4} \mu^z} 
    e^{i\frac{\pi}{4} \mu^x \tau^x} f_q$  
  \\ \hline 
  $ F_{0,q}^\dagger \rightarrow$ & $-F_{0,q}^\dagger$ & 
  $-F_{0,q}^\dagger$ & $iF_{0,q}^\dagger$ & $F_{0,q}^\dagger$ & 
  $F_{0,-q}$ & $F_{0,-q}^\dagger$ & $-e^{iq\pi/3}F_{0,q}^\dagger$ 
  \\ \hline 
  $ F_{1,q}^\dagger \rightarrow$ & $F_{1,q}^\dagger$ & 
  $F_{1,q}^\dagger$ & $iF_{2,q}^\dagger$ & $-F_{1,q}^\dagger$ & 
  $F_{1,-q}$ & $F_{1,-q}^\dagger$ & $e^{iq\pi/3}F_{3,q}^\dagger$ 
  \\ \hline 
  $ F_{2,q}^\dagger \rightarrow$ & $F_{2,q}^\dagger$ & 
  $-F_{2,q}^\dagger$ & $iF_{1,q}^\dagger$ & $-F_{2,q}^\dagger$ & 
  $F_{2,-q}$ & $F_{2,-q}^\dagger$ & $e^{iq\pi/3}F_{1,q}^\dagger$  
  \\ \hline 
  $ F_{3,q}^\dagger \rightarrow$ & $-F_{3,q}^\dagger$ & 
  $F_{3,q}^\dagger$ & $iF_{3,q}^\dagger$ & $-F_{3,q}^\dagger$ & 
  $F_{3,-q}$ & $F_{3,-q}^\dagger$ & $e^{iq\pi/3}F_{2,q}^\dagger$ 
  \\ \hline 
  $ F_{R,q}^\dagger \rightarrow$ 
    & $e^{-2i\tilde Q}F_{R,q}^\dagger$  
  & $e^{i(\tilde Q-\frac{\pi}{2})}F_{R,q}^\dagger$ &  
  $iF_{R,q}^\dagger$ & $F_{L,q}^\dagger$ & $-F_{L,-q}$ & 
  $F_{L,-q}^\dagger$ & $-e^{iq\pi/3}F_{L,q}^\dagger$ 
  \\ \hline 
  $ F_{L,q}^\dagger \rightarrow$ 
    & $e^{2i\tilde Q}F_{L,q}^\dagger$  
  & $e^{-i(\tilde Q-\frac{\pi}{2})}F_{L,q}^\dagger$ & 
  $iF_{L,q}^\dagger$ & $F_{R,q}^\dagger$ & $-F_{R,-q}$ & 
  $F_{R,-q}^\dagger$ & $-e^{iq\pi/3}F_{R,q}^\dagger$ 
\end{tabular}  
\end{ruledtabular}  
\end{table*}  
 
We now introduce the monopole operators by specifying 
their action on the ground state with no added flux, denoted 
$|0\rangle$.  First, we define monopole creation operators 
$M_{\alpha}^\dagger$ which insert $+2\pi$ flux and fill the zero-modes as 
follows, 
\begin{eqnarray} 
  M^\dagger_0 |0\rangle &=& e^{i\alpha_0}F^\dagger_{0,+1}|DS,+1\rangle 
  \label{M0} \\ 
  M^\dagger_j |0\rangle &=& e^{i\alpha_j}F^\dagger_{j,+1}|DS,+1\rangle 
  \label{Mj} \\ 
  M^\dagger_{R/L} |0\rangle &=& 
  e^{i\alpha_{R/L}}F^\dagger_{R/L,+1}|DS,+1\rangle. 
  \label{Mrl} 
\end{eqnarray} 
Here $j$ runs from 1 to 3 and $|DS,+1\rangle$ is the filled 
negative-energy Dirac sea with 
$+2\pi$ flux inserted and all zero modes vacant.  The Hermitian 
conjugate operators $M_\alpha$ are required to add the opposite 
momentum and flux to the ground 
state, 
\begin{eqnarray} 
  M_0 |0\rangle &=& e^{i\beta_0}F^\dagger_{0,-1}|DS,-1\rangle 
  \label{antiM0} \\ 
  M_j |0\rangle &=& e^{i\beta_j}F^\dagger_{j,-1}|DS,-1\rangle 
  \label{antiMj} \\ 
  M_{R/L} |0\rangle &=& 
  e^{i\beta_{R/L}}F^\dagger_{L/R,-1}|DS,-1\rangle, 
  \label{antiMrl} 
\end{eqnarray} 
where $|DS,-\rangle$ is the filled negative-energy Dirac sea with  
$-2\pi$ flux inserted.  It 
is important to note that the phases $\alpha$ and $\beta$ in the 
definitions above  
are arbitrary, and can be specified to our convenience so as to 
construct operators with the desired transformation properties.   
 
To obtain a continuum expression for $S^+$ in terms of the monopole 
operators, we need to determine the quantum numbers they carry.  This 
in turn requires knowing the transformation properties of 
$|DS,q\rangle\langle0|$.  We take up this rather involved issue in 
Appendix \ref{MonopoleApp}.  By employing general relations among the 
symmetries (such as $R_\pi^2 = 1$, {\it etc}.), we first establish that 
\begin{eqnarray} 
  T_{1,2} &:& |DS,q\rangle\langle 0| \rightarrow -|DS,q\rangle\langle 0| 
  \nonumber \\ 
  \tilde {\mathcal R}_x &:& |DS,q\rangle\langle 0| \rightarrow  
  i\zeta_x |DS,q\rangle\langle 0| 
  \nonumber \\ 
  R_\pi &:& |DS,q\rangle\langle 0| \rightarrow  
  \zeta_{\pi} |DS,q\rangle\langle 0| 
  \nonumber \\ 
  {\mathcal C} &:& |DS,q\rangle\langle 0| \rightarrow  
  \prod_{Aa} 
  f^\dagger_{Aa,-q}|DS,-q\rangle\langle 0|  
  \nonumber \\ 
  {\mathcal T}_{\rm ferm} &:& |DS,q\rangle\langle 0| \rightarrow 
  |DS,-q\rangle\langle 0| 
  \nonumber \\ 
  R_{\pi/3} &:& |DS,q\rangle\langle 0| \rightarrow 
  \zeta_{\pi} e^{2iq\pi/3}|DS,q\rangle\langle 0|. 
  \label{DiracSeaSymmetry} 
\end{eqnarray}  
Here $\zeta_{x}$ and $\zeta_{\pi}$ are signs which in principle 
are fixed, but can not be determined using only symmetry relations.   
The specified action under rotation applies only in the isotropic 
limit.   
 
\begin{figure}  
  \begin{center}  
    {\resizebox{6cm}{!}{\includegraphics{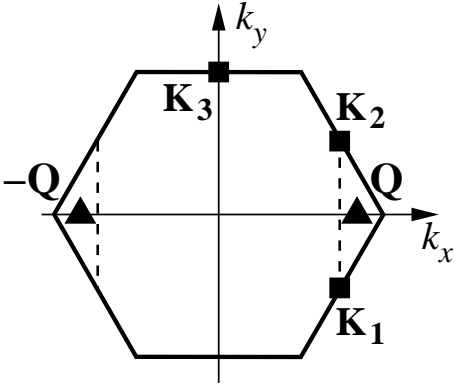}}}  
  \end{center}  
  \caption{Momenta at which the dynamical spin structure factor 
    $S^{+-}$ exhibits dominant power law correlations with the  
    \emph{same exponent} in the AVL.  The leading power law in the 
    $S^{zz}$ structure factor, by contrast, occurs only at momenta 
    $\pm {\bf Q}$.}  
  \label{MonopoleQs}  
\end{figure}

This information is sufficient to determine the momenta carried by the 
monopoles: $M_0$ carries zero momentum, $M_j$ carries momentum  
${\bf K}_j$ on the midpoint of a Brillouin zone edge  
(cf.~Fig.~\ref{MonopoleQs}), and $M_{R/L}$  
carry momenta at the spiral ordering 
wave vectors $\pm{\bf Q}$.  For other symmetries, however, a more 
careful analysis is required.  In fact, by examining the symmetry 
of the monopoles under inversion, one can show that it is 
\emph{impossible} for all six bare monopole operators to contribute to $S^+$.   
Under inversion, we have $M_0 \rightarrow \zeta_{\pi} 
M_0$ while $M_j \rightarrow -\zeta_{\pi} M_j$.  But the Fourier 
components of $S^+$ at zero momentum and ${\bf K}_j$ are all 
\emph{even} under inversion.  Depending on the sign  
$\zeta_{\pi}$, either the bare monopole $M_0$ or the three monopoles $M_j$ 
must therefore be excluded on symmetry grounds from a continuum  
expression for $S^+$.  (Since $M_{R/L}$ are not diagonal under 
inversion, one can always choose the phases $\alpha_{R/L}$ and 
$\beta_{R/L}$ to construct operators that transform like $S^+$ at momenta  
$\pm {\bf Q}$).   
 
To determine the remaining ambiguities we appeal to numerical studies 
of monopole insertions.  
Specifically, we diagonalize the mean-field hopping Hamiltonian on a 
finite system with arbitrary flux insertions to obtain the 
single-particle energies and wave functions.  With these wave 
functions in hand, it is then possible to obtain the inversion and  
reflection properties of $|DS,q\rangle\langle 0|$.   
These numerics are discussed in more detail in Appendix 
\ref{MonopoleApp}.  To ensure geometry-independence of the results, a 
variety of boundary conditions and system sizes were considered.  In 
all cases, we find that $\zeta_{\pi} = \zeta_x = -1$.  In particular, 
we conclude 
that $M_0$ by itself does not contribute to a continuum expression for 
$S^+$.   
 
We now have enough information to determine unambiguously all quantum 
numbers carried by the monopoles.  It is straightforward to show that 
the phases appearing in Eqs.\ (\ref{M0}) through (\ref{antiMrl}) can 
be chosen so that the monopoles transform as shown in Table 
\ref{tab:monopoles}.  (Note that under physical time-reversal 
$S^+\rightarrow -S^-$, whereas ${\mathcal T}_{\rm ferm}$ sends 
$M_{\alpha}^\dagger \rightarrow +M_{\alpha}$.  This is not too 
surprising, however, given that the U(1) spin symmetry is not manifest 
in the dual theory.)   
The desired continuum expression for $S^+$ can then be written as  
follows, 
\begin{eqnarray} 
  S^+ \sim [e^{-i{\bf Q}\cdot{\bf r}}M_R^\dagger +  
  e^{i{\bf Q}\cdot{\bf r}}M_L^\dagger]  
  + \sum_{j = 1}^3 e^{i{\bf K}_j\cdot {\bf r}}M_j^\dagger 
  +\cdots, 
\end{eqnarray} 
where the ellipsis represents subdominant contributions.  The momenta 
$\pm{\bf Q}$ and ${\bf K}_j$ carried by the monopoles on the 
right-hand-side are sketched in Fig.\ \ref{MonopoleQs}. 
 
Monopole operators are known to have nontrivial power-law correlations 
in large-$N$ QED3, each with \emph{identical} scaling dimension  
$\Delta_{\rm m} \approx 0.26N$.\cite{BKW}  This fact leads us to a 
nontrivial prediction for the in-plane spin structure factor 
$S^{+-}$ in the AVL.  Namely, $S^{+-}$ exhibits the same universal 
power-law correlations at each of the five momenta 
${\boldsymbol \Pi}_j$ of Fig.\ \ref{MonopoleQs}.   This remarkable 
property stems from the enlarged global SU(4) flavor symmetry enjoyed by 
the AVL.    
For wave vectors near ${\boldsymbol \Pi}_j$, the in-plane structure 
factor thus scales as 
\begin{equation} 
  S^{+-}({\bf k} = {\boldsymbol \Pi}_j+{\bf q},\omega) \sim  
  A_{j} \frac{\Theta(\omega^2-{\bf q}^2)}{(\omega^2-{\bf q}^2)^{1-\eta_{\rm 
  m}/2}}. 
\end{equation} 
The anomalous dimension $\eta_{\rm m}$ is given in the large-$N$ limit 
by 
\begin{equation} 
  \eta_{\rm m} \approx 0.53 N - 1. 
\end{equation} 
Setting $N = 4$ yields $\eta_{\rm m} \approx 1.12$ and $\Delta_m 
\approx 1.04$.  While the scaling dimension is the same at each wave  
vector, it is important to keep in mind that the amplitudes $A_j$ can vary  
significantly at different momenta.  For example, near the limit of 
decoupled chains the amplitude at ${\bf K}_3$ should be much 
suppressed relative to the other four wave vectors, which are near 
$k_x = \pi$ where most of the activity would be expected.   
   
We note here that while exclusion of the bare monopole $M_0$ from a 
continuum expression for $S^+$ was not obvious at the outset, this 
conclusion is quite reasonable physically in light of the spin 
correlations discussed above.  If this exclusion did  
\emph{not} occur, then the dynamic spin structure factor $S^{+-}$ would 
exhibit the same power-law correlations at zero momentum and the five 
wave vectors in Fig.\ \ref{MonopoleQs}.  This would be quite 
surprising given that one would intuitively expect subdominant 
correlations at zero momentum in an antiferromagnet.     
 
The locations of the leading in-plane correlations in the AVL  
are suggestive of proximity to magnetically ordered phases 
involving condensation of $S^+$ at the momenta shown in Fig.\ 
\ref{MonopoleQs}.  We will explore some of these states below.

\begin{table*} 
\caption{\label{tab:monopoles} Transformation properties of 
the six monopole operators in QED3.  }  
\begin{ruledtabular}  
\begin{tabular}{c | c | c | c | c | c | c}  
  & $T_{\delta{\bf r}}$ & $\tilde {\cal R}_x$ & $R_\pi$  
  & ${\mathcal C}$ & ${\cal T}_{\rm ferm}$ & $R_{\pi/3}$ (isotropic limit) 
  \\ \hline 
  $M_{0} \to$ & $M_{0}$ & $M_{0}$ & $-M_{0}$ & $-M_{0}^\dagger$ &  
  $M_{0}^\dagger$ & $-M_{0}$  
  \\ \hline 
  $M_{1} \to$ & $e^{i{\bf K}_1\cdot{\bf \delta r}}M_{1}$ & $M_{2}$  
  & $M_{1}$ & $M_{1}^\dagger$ & $M_{1}^\dagger$ & $M_{3}$  
  \\ \hline 
  $M_{2} \to$ & $e^{i{\bf K}_2\cdot{\bf \delta r}}M_{2}$ & $M_{1}$  
  & $M_{2}$ & $M_{2}^\dagger$ & $M_{2}^\dagger$ & $M_{1}$  
  \\ \hline 
  $M_{3} \to$ & $e^{i{\bf K}_3\cdot{\bf \delta r}}M_{3}$ & $M_{3}$  
  & $M_{3}$ & $M_{3}^\dagger$ & $M_{3}^\dagger$ & $M_{2}$  
  \\ \hline 
  $M_{R/L} \to$ & $e^{\pm i{\bf Q}\cdot{\bf \delta r}}M_{R/L}$ & $M_{R/L}$  
  & $M_{L/R}$ & $M_{L/R}^\dagger$ & $M_{R/L}^\dagger$ & $M_{L/R}$  
\end{tabular}  
\end{ruledtabular}  
\end{table*}

\section{Proximate Phases to the Algebraic Vortex Liquid} 
\label{NearbyPhases} 
 
In this section we explore the neighboring phases of the AVL  
that are encoded by the effective Lagrangian (\ref{L0}).   
Response properties of the bosonic spin system will be obtained by 
introducing an external probing field $A^{\rm ext}$ that couples  
in the dual theory via Eq.\ (\ref{deltaL}).  As discussed in Sec.\ 
\ref{AVLstability}, we will assume 
$\kappa = 0$ so that $A^{\rm ext}$ couples to both the original boson 
currents $\delta j = (\nabla \times a)/2\pi$ and the Chern-Simons flux.   
This exploration will provide some guidance as to where in the phase diagram  
the AVL lies, and is also useful because a study of the 
phase diagram within a direct Landau analysis of the spin model is 
hindered by Berry phases.  
 
Descendants of the AVL are obtained by giving the fermions a mass, 
which destabilizes the vortex liquid leading to a wealth of possible states.   
Here, we will restrict our attention to nearby states favored by the  
interactions ${\mathcal L}_{\rm 4f}$.  Although we postulated above that 
such terms are irrelevant in the AVL critical theory, sufficiently strong 
four-fermion interactions are nevertheless expected to generate fermion 
masses and destroy the AVL.  We will focus, in particular, on states  
arising from the generation of \emph{enhanced} fermion mass terms,  
as the AVL will 
likely be more susceptible to realizing such states.  Moreover, for 
simplicity we will consider the spatially isotropic limit $J = J'$.   
In this case, the four-fermion terms  
${\mathcal L}_{\rm 4f}$ can be written in terms of enhanced bilinears 
in Table \ref{tab:EnhancedBilinears} as follows, 
\begin{eqnarray} 
  {\mathcal L}_{\rm{4f}} &=& u_1 {\mathcal M}_{KL}^2  
  + u_2 {\mathcal M}_{\sqrt{3}\times\sqrt{3}}^2 + 
  u_3 {\mathcal M}_{SS}^\dagger {\mathcal M}_{SS}  
  \nonumber \\ 
  &+& \sum_{j = 1}^3 (u_4 {\mathcal K}_j^2 
  + u_5 {\mathcal K}^{\prime 2}_j+u_6 {\mathcal P}_j^\dagger {\mathcal P}_j)  
  +{\mathcal L}_{\rm 4f,ne}. 
  \label{L4f} 
\end{eqnarray}   
The last term ${\mathcal L}_{\rm 4f,ne}$ represents four-fermion interactions 
composed entirely of non-enhanced bilinears, which will not be of 
interest here.  Though not unique,  
this form provides a useful organization of 
the four-fermion interactions based on their translation and rotation 
properties.  Our exploration below is by no means 
intended to be exhaustive; rather, our aim is to illustrate some  
representative  
examples of the proximate phases that can be analyzed in the 
fermionized vortex theory.

\subsection{Kalmeyer-Laughlin spin liquid} 
\label{KLsl} 
 
Consider first the addition of a mass term  
$m {\mathcal M}_{KL}=m\barpsi \psi$,  
which is favored by a large negative $u_1$ interaction above.   
This mass respects all symmetries except ${\mathcal T}_{\rm ferm}$,  
and drives the system  
into a $\nu = 1/2$ fractional quantum 
Hall state for the original bosons,  
which breaks \emph{physical} time-reversal symmetry.   
In other words, the physical spin state obtained by the addition of  
$m {\mathcal M}_{KL}$ is  
the Kalmeyer-Laughlin chiral spin-liquid.\cite{KLshort,KLlong}   
 
To demonstrate this, let's first integrate out the massive fermions.   
Since all flavors have 
the same mass $m$ with the same sign, integrating out the fermions 
induces a Chern-Simons term for $(a+A)_\mu$.  The Lagrangian 
is then 
\begin{eqnarray} 
  {\cal L}_{a, A} &=&   
  \frac{1}{2 e^2} (\nabla \times a)^2 +  
  \frac{i}{4\pi} A \cdot (\nabla \times A) 
  \nonumber \\  
  && + \frac{i \text{sign}(m)}{2\pi} (a+A) \cdot [\nabla \times (a+A)] 
  \nonumber \\  
  && -\frac{i}{2\pi}A^{\rm ext} \cdot [\nabla \times (a+A)] ~. 
\end{eqnarray} 
The spectrum for the 
above Lagrangian is gapped, which can be verified by integrating out 
the Chern-Simons field $A$.  Integrating out further 
the gauge field $a$, we arrive at an effective Lagrangian for the 
probing field $A^{\rm ext}$, 
\begin{equation} 
  {\mathcal L}_{\rm eff} =  
  -\text{sign}(m)\frac{i\sigma_{xy}}{2}A^{\rm ext}\cdot 
  (\nabla \times A^{\rm ext}), 
  \label{Lext} 
\end{equation} 
with $\sigma_{xy} = \nu/2\pi = 1/4\pi$.   
Thus Eq.\ (\ref{Lext}) characterizes the response for a  
$\nu = 1/2$ fractional quantum Hall 
state of the original bosons as claimed. 
We remark here that this physics results with either sign for the mass 
$m$.  Had we instead chosen to couple $A^{\rm ext}$ only to the 
original boson currents [\emph{i.e.}, with $\kappa = 1$ in Eq.\ 
(\ref{deltaL})], then  
a particular sign of the mass would have to be chosen relative to the sign of 
the Chern-Simons flux in order to recover the Kalmeyer-Laughlin state.   
Once again, we see that endowing the 
Chern-Simons flux with boson charge leads to response properties of 
the spin system that are insensitive to the direction of flux 
attachment as desired.   
  
What is the nature of the gapped 
excitations in this phase?  Consider acting on the ground state with the 
fermion field $\psi^\dagger$.  The added fermion couples to the 
sum $\tilde{a} = a+A$.  By examining the action obtained by retaining 
$\tilde a$ and integrating out the Chern-Simons field,  
we see that the system dynamics binds $\Delta\times\tilde{a} = -\pi$  
flux to the fermion (which also carries $2\pi$ Chern-Simons flux).   
Thus, the fermion is turned into a 
\emph{semionic} excitation carrying spin-1/2.  This is precisely the 
gapped semionic spinon in the Kalmeyer-Laughlin state.

\subsection{Magnetically ordered phases} 
\label{MagOrdPhases} 
The remaining states we consider arise from generating specific 
fermion mass terms of the form $m \barpsi \hat W \psi$, where $\hat W$ 
has two $+1$ and two $-1$ eigenvalues.  Hence, two fermion modes have mass 
$m$, while the other two have mass $-m$.  In all such phases, the 
vortices are ``insulating,'' and the ``photon'' in the dual 
gauge field $a$ can freely propagate.  The gapless 
photon is revealed upon 
integrating out the massive fermions, which induces only a 
generic Maxwell term for the field $\tilde{a}=a+A$.  These vortex 
insulators correspond to magnetically ordered phases of the spin 
system.  The gapless photon is the Goldstone spin-wave at 
zero-momentum arising from the broken continuous U(1) spin symmetry. 
Moreover, the probing field $A^{\rm ext}$ is massive here, which is 
the ``Meissner effect'' expected for the superfluid phase of the 
original bosons. 
Our objective below will be to disentangle the spin order that  
arises in different vortex insulators.  
As we will see, magnetically ordered states neighboring the AVL 
fall into two categories: 
conventional XY spin-ordered phases and ``supersolids,'' which 
additionally develop $S^z$ order.

\subsubsection{XY spin-ordered states} 
\label{XYorder} 
 
\begin{figure}  
  \begin{center}  
    {\resizebox{7.5cm}{!}{\includegraphics{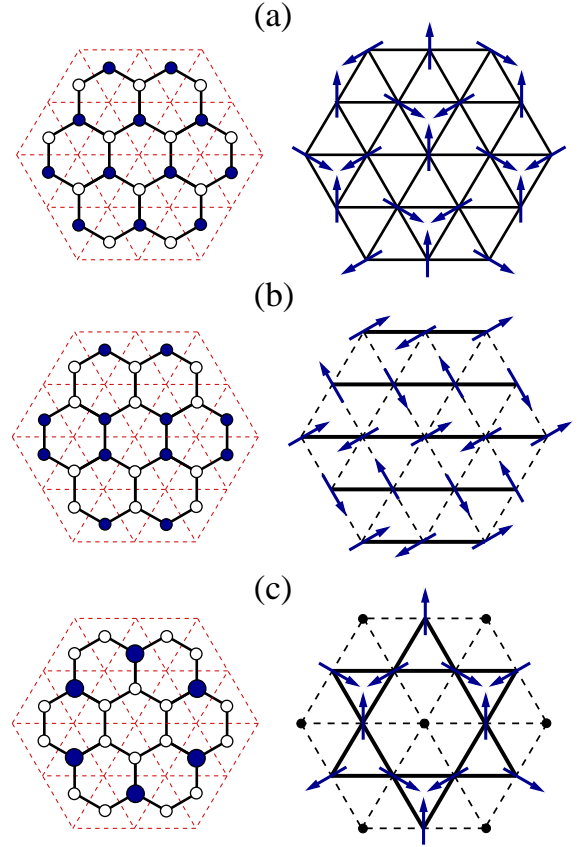}}}  
  \end{center}  
  \caption{ 
  Vortex charge density waves (CDWs) proximate to the AVL in the
  isotropic limit $J = J'$, along with the corresponding spin 
  structures.  Vortices preferentially occupy 
  filled honeycomb sites.  On the right side the satisfied bonds of 
  the triangular lattice are solid, dashed lines represent less 
  satisfied bonds, and filled circles denote sites whose spins 
  fluctuate around zero mean.}  
  \label{CDWs}  
\end{figure}

Consider the addition of a mass term $m {\mathcal M}_{\sqrt{3}\times\sqrt{3}}$, 
which is favored by a large negative $u_2$ in Eq.\ (\ref{L4f}). 
Microscopically, this mass can be identified with  
a staggered vortex chemical potential that causes the vortices to 
preferentially occupy one of the two sublattices of the honeycomb.   
The resulting state is the vortex ``charge density wave'' (CDW) shown in 
Fig.\ \ref{CDWs}(a), where the vortex density is enhanced on the 
filled sites and depleted on the open sites.   
 
To identify the corresponding spin structure,  
first recall that the leading $S^{+-}$ spin 
correlations in the AVL occur at wave vectors  
$\pm {\bf Q}$ and ${\bf K}_{1,2,3}$ of Fig.\ \ref{MonopoleQs}.  It is 
natural, then, to expect that magnetically ordered descendants of the 
AVL will involve condensation of $S^{+}$ at these wave vectors.   
We will assume this is the case, and search for the symmetry-equivalent of 
${\mathcal M}_{\sqrt{3}\times\sqrt{3}}$ by considering bilinears involving 
$S^{\pm}$ at these wave vectors.  The answer is unique (up to an overall 
sign), and we identify  
\begin{equation} 
  {\mathcal M}_{\sqrt{3}\times\sqrt{3}}  
  \sim S^+_{\bf Q}S^-_{\bf Q}-S^+_{\bf -Q}S^-_{\bf -Q}.  
  \label{SpiralOP} 
\end{equation}  
Since $\langle{\mathcal M}_{\sqrt{3}\times\sqrt{3}}\rangle \neq 0$ in 
this vortex CDW, it follows that the spin order can be obtained from  
$\langle S^+_{\bf Q}\rangle \neq 0$,  
$\langle S^+_{\bf -Q}\rangle = 0$ (or vice versa, depending on the 
sign of the mass $m$).  This is the well known  
$\sqrt{3} \times \sqrt{3}$ spiral state depicted in Fig.\ \ref{CDWs}(a).  We 
note that our identification in Eq.\ (\ref{SpiralOP}) holds in the 
anisotropic case as well; in this regime an incommensurate spiral 
results. 
 
As another example, assume the $u_4$ interaction is strong enough that a 
mass term $\sum_j m_j {\mathcal K}_j$ is generated.  This mass similarly 
corresponds to a modulated vortex chemical potential which drives CDW 
ordering.  With only 
quartic fermion terms, there is a large degeneracy of possible states  
due to the 
arbitrariness in the relative values of $m_{1,2,3}$.  This degeneracy is  
broken, however, by higher-order terms in the action, which select 
either (I) $m_i \neq 0$, $m_{j\neq i} = 0$, or (II)  
$|m_1| = |m_2| = |m_3| \neq 0$.   
The resulting vortex CDW's are shown in Figs.\ \ref{CDWs}(b) and (c), 
respectively.   
 
The spin order in these states can be determined using 
the same logic as above.  Here, we identify 
\begin{equation} 
  {\mathcal K}_i \sim i \epsilon_{ijk}S^+_{{\bf K}_j}S^-_{{\bf K}_k}.  
  \label{Kidentification} 
\end{equation} 
Consider case (I) first, with say $\langle{\mathcal K}_{1,2}\rangle = 0$ 
and $\langle {\mathcal K}_{3}\rangle \neq 0$.  Equation 
(\ref{Kidentification}) then implies that  
$\langle S^+_{{\bf K}_3}\rangle = 0$ and 
$\langle S^+_{{\bf K}_{1,2}}\rangle \sim e^{i\varphi_{1,2}}$, which 
yields $\langle {\mathcal K}_{3}\rangle \sim \sin(\varphi_1-\varphi_2)$.   
To maximize $|\langle {\mathcal K}_3\rangle|$, which is energetically  
favored by the large $u_4$ interaction, the phases $\varphi_1$ and  
$\varphi_2$ are chosen to differ by $\pi/2$.  The resulting spin order 
is shown on the right-hand-side of Fig.\ \ref{CDWs}(b).  Now let's  
consider case (II), where $|\langle {\mathcal K}_1\rangle| = |\langle 
{\mathcal K}_2\rangle| = |\langle {\mathcal K}_3\rangle|\neq 0$.   
Here we take $\langle S^{+}_{{\bf K}_j}\rangle \sim e^{i\varphi_j}$, 
yielding $\langle {\mathcal K}_1\rangle \sim \sin(\varphi_2-\varphi_3)$, 
\emph{etc}.  In this case the phases $\varphi_{1,2,3}$ must differ by 
either $\pi/3$ or $2\pi/3$, which leads to the spin order  
illustrated in Fig.\ \ref{CDWs}(c).     
On the right-hand-side of Fig.\ \ref{CDWs}, 
the solid lines indicate satisfied bonds, while the filled circles 
denote triangular lattice sites with spins fluctuating around zero 
mean.   
 
A similar analysis can be used to identify the states arising from 
spontaneously generated mass involving the enhanced bilinears ${\mathcal P}_j$. 
Such mass terms give rise to modulated nearest-neighbor hopping  
amplitudes for the 
vortices, and drive vortex ``valence bond solid'' (VBS) order.  Since 
${\mathcal P}_j$ carries momentum ${\bf K}_j+{\bf Q}$, the spin structures 
corresponding to these VBS phases involve condensation of both 
$S^{+}_{\bf Q}$ and $S^+_{{\bf K}_j}$.  At present it is unclear what 
order is driven by ${\mathcal K}'_j$ mass terms, due to the fact that 
they arise from second-neighbor vortex hopping, which does not have a 
clear interpretation for the spin system.

\subsubsection{Supersolids} 
\label{SSphases} 
 
\begin{figure}  
  \begin{center}  
    {\resizebox{7.5cm}{!}{\includegraphics{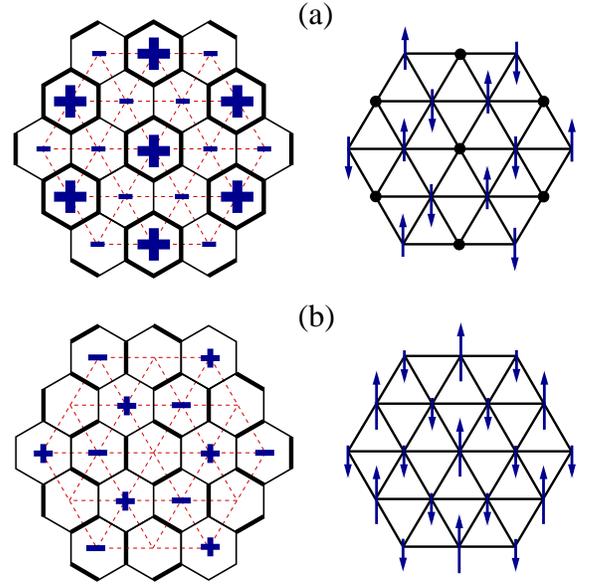}}}  
  \end{center}  
  \caption{Supersolids neighboring the AVL.  On the left side, the 
  direction of induced gauge flux is specified by a $\pm$ sign, and 
  the hopping amplitudes are dominant along bold links of the 
  honeycomb.  The ``solid'' ordering pattern for  
  $\langle S^z\rangle$ follows the pattern of induced gauge flux, 
  while the ``superfluid'' order for $\langle S^+\rangle$ is shown on 
  the right side.  Filled circles denote sites with $\langle 
  S^+\rangle = 0$.  The overall spin structure can be viewed as a  
  spiral state, tilted into the $(S^y,S^z)$ plane. 
  }  
  \label{SuperSolid}  
\end{figure}

Finally, consider a mass term  
$m [e^{i\gamma} M^\dagger_{SS}+{\rm H.c.}]$ generated by a  
large $u_3$ interaction.  Microscopically, this mass 
induces both modulations in the nearest-neighbor vortex hopping 
amplitudes and modulations in the gauge flux piercing the honeycomb 
plaquettes.  The degeneracy in the phase $\gamma$ is lifted by 
higher-order terms in the action, which select either $\gamma = 
n\pi/3$ or $\gamma = (2n+1)\pi/6$, where $n$ is an integer. 
The vortex states for these two cases are shown in Figs.\ 
\ref{SuperSolid}(a) and (b), respectively.  In the figure, the 
direction of induced flux through a given plaquette is indicated with 
a $\pm$ sign, and the bold honeycomb links have the  
dominant hopping amplitudes.  In Fig.\ \ref{SuperSolid}(a), the 
induced flux is twice as large on ``$+$'' plaquettes; in (b), the 
induced flux is equal and opposite on ``$\pm$'' plaquettes and 
vanishes on others. 
  
These lattice-scale gauge flux modulations signify the onset 
of $S^z$ ordering in the spin system, 
\begin{equation} 
  \langle S^z\rangle \sim \cos({\bf Q\cdot r}+\gamma). 
\end{equation} 
Since there is a gapless photon in these states, the 
continuous U(1) spin symmetry is also broken.  Hence 
the in-plane spin components order as well, so that these states are 
examples of ``supersolids''.  Using the symmetry of the vortex phases 
and the uncertainty principle as a guide, the simplest assumption for the  
$S^+$ order is shown on the right side 
of Figs.\ \ref{SuperSolid}(a) and (b).  The filled circles in the 
figure denote 
sites with $\langle S^+\rangle = 0$.  The in-plane and out-of-plane 
spin structure can be collectively viewed as a coplanar spiral state 
rotated into, for instance, the $(S^y,S^z)$ plane. 
Both spin patterns exhibit a $\sqrt{3} \times \sqrt{3}$ periodicity. 
The difference is that in Fig.\ \ref{SuperSolid}(a) spins on one 
sublattice point along the hard $S^z$ axis, while in Fig.\ \ref{SuperSolid}(b) 
spins on one sublattice point along the $S^y$ axis.

\section{Discussion} 
\label{Conclusion} 
 
In this paper we have performed a detailed characterization of a novel 
``critical'' spin liquid, the  
algebraic vortex liquid, which arises rather naturally out of a 
reformulation of the easy-plane spin model in terms of fermionized vortex  
degrees of freedom. 
Among the most striking predictions for the AVL is 
the behavior of the dynamical spin correlations.  As a consequence 
of an emergent global SU(4) symmetry, the in-plane spin structure factor 
$S^{+-}$ exhibits enhanced power law correlations with 
\emph{identical} exponents at the five inequivalent momenta shown in Fig.\ 
\ref{MonopoleQs}.  Due to the easy-plane character of the AVL, the  
out-of-plane spin correlations behave quite differently.  The $S^{zz}$ 
structure factor exhibits enhanced 
power law correlations only at the spiral ordering wave vectors 
($\pm {\bf Q}$ in Fig.\ \ref{MonopoleQs}), and is generally  
expected to be weaker than $S^{+-}$.  These nontrivial features in the 
spin structure factor 
distinguish the AVL from other known spin liquids, and should serve as 
useful characterizations for identifying this phase experimentally. 
 
Our study was partly motivated by the spin-1/2 triangular antiferromagnet 
\CsCuCl, whose 
spin dynamics have been explored with neutron 
scattering.\cite{ColdeaShort,ColdeaLong}  Although this material develops 
long range spiral order at low temperatures $T \lesssim T_N = 0.62 $K, the 
dynamical structure factor exhibits anomalous power laws at 
intermediate energies, both in the ordered phase and in a range of 
temperatures above $T_N$.  Such power law behavior is reminiscent of 
spin-liquid physics, and several scenarios for its origin have been 
proposed.   
These include physics dominated by one-dimensional 
chains,\cite{Bocquet, Starykh} 
a two-dimensional algebraic spin liquid,\cite{ZhouWen}  
the $Z_2$ spin liquid,\cite{Z2anisotropic, SpinonStatistics} and a 
quantum critical point scenario,\cite{Isakov} as well as more  
conventional explanations such as nonlinear spin wave theory.\cite{Veillette2} 
 
Possible application of the 
AVL to \CsCuCl~has been discussed in some detail in Ref.\ 
\onlinecite{AVLshort}.  The most intriguing observation 
here is that the dynamical  
structure factor was found experimentally to decay with the same power  
law near wave vectors ${\bf K}_{1,2}$ and ${\bf Q}$ in Fig.\ 
\ref{MonopoleQs},\cite{ColdeaLong} consistent with our expectations 
for the AVL.  
It is important to keep in mind, however, that  
the dual vortex formulation employed here requires an easy-plane 
U(1) spin symmetry so that vortices exist as stable topological 
excitations.  This imparts the AVL with a distinct easy-plane 
character, unlike other theoretical proposals which retain full SU(2) 
spin symmetry.  Although there is a microscopic easy-plane spin 
anisotropy in \CsCuCl\ due to a \DM\ interaction, this coupling is 
fairly weak, and it is therefore not clear whether the AVL described 
here can be applied directly  
(some scenarios are considered in Ref.~\onlinecite{AVLshort}). 
An interesting possibility is that the AVL has an SU(2)-invariant  
relative which may be relevant for \CsCuCl, though we do not know how  
to access such a state theoretically. 
One speculation in this respect is that there may exist a slave-particle 
description of the AVL.  In particular, the direct slave-fermion approach 
often yields critical states dubbed algebraic spin liquids (ASLs); 
such states on the triangular lattice were explored in  
Ref.~\onlinecite{ZhouWen}. 
The algebraic vortex liquid is not formulated using spinon fields, 
but predicts critical power law spin correlations reminiscent of those 
in ASLs.  It should also be said that spinons are strongly interacting 
in the ASL and cannot be thought of as free fields in any sense, 
and the same is true about vortices in the AVL. 
Unfortunately, so far we have been unable to find a spin liquid 
state on the triangular lattice that would reproduce all the dominant 
wave vectors present in the AVL phase, but such a connection 
between these very different theoretical perspectives remains 
a tantalizing possibility.

\begin{acknowledgments}  
 
We would like to acknowledge Leon Balents and T.\ Senthil for many  
illuminating discussions, and especially Mike Hermele for an initial  
collaboration. 
This work was supported by the National Science Foundation (J.\ A.) 
through grants PHY-9907949 (O.~I.~M.\ and M.\ P.\ A.\ F.) and  
DMR-0529399 (M.\ P.\ A.\ F.).    
  
\end{acknowledgments}

\appendix 

\section{Transformation properties of the negative-energy Dirac sea} 
\label{MonopoleApp} 
 
This appendix is devoted to obtaining the transformation properties of 
$|DS,q\rangle\langle 0|$, which are needed for determining the 
quantum numbers carried by the monopole operators in QED3.   
Here $|DS,q\rangle $ is the filled 
negative-energy Dirac sea with a $2\pi q$ flux insertion, where $q = 
\pm 1$, and $|0\rangle$ is the ground state with no added flux.  We 
attack the problem in two stages.  First, we constrain the 
transformation properties as much as possible using various general  
relations among symmetries.  The ambiguities that still remain here are then 
fixed using numerical studies of monopole insertions.  
 
\subsubsection{General arguments}

{\it Fermionic time reversal and Particle-hole symmetry}.  By 
examining Table \ref{tab:d}, we see that the flux 
changes sign under both ${\mathcal T}_{\rm ferm}$ and  
${\mathcal C}$.  Hence ${\mathcal T}_{\rm ferm}$ transforms the filled 
negative-energy Dirac sea with $q = +1$ into the negative-energy Dirac 
sea with $q = -1$, while ${\mathcal C}$ additionally fills the four 
zero-modes since $|DS,q\rangle$ is not half-filled.   
The ground state $|0\rangle$, on the other hand, is an 
eigenstate of both symmetries.  Using ${\mathcal T}_{\rm ferm}^2 = 
{\mathcal C}^2 = 1$, we can then define the phases of $|DS,q\rangle$ 
such that  
\begin{eqnarray} 
  {\mathcal T}_{\rm ferm} &:& |DS,q\rangle\langle0| \rightarrow 
  |DS,-q\rangle \langle 0| 
  \\ 
  {\mathcal C} &:& |DS,q\rangle\langle 0| \rightarrow \prod_{Aa} 
  f_{Aa,-q}^\dagger |DS,-q\rangle \langle 0|. 
\end{eqnarray} 
 
Both $|DS,q\rangle$ and $|0\rangle$ are expected to be eigenstates 
under the remaining symmetries in Table \ref{tab:f}, all of which 
leave the flux invariant.  Quite generally, we then have 
\begin{eqnarray} 
  R_\pi &:& |DS,q\rangle \langle 0| \rightarrow e^{i\theta_\pi^q} 
  |DS,q\rangle\langle 0| 
  \nonumber \\ 
  \tilde {\mathcal R}_x &:& |DS,q\rangle \langle 0| \rightarrow e^{i\theta_x^q} 
  |DS,q\rangle\langle 0| 
  \nonumber \\ 
  T_{1,2} &:& |DS,q\rangle \langle 0| \rightarrow e^{i\theta_{1,2}^q} 
  |DS,q\rangle\langle 0| 
  \nonumber \\ 
  R_{\pi/3} &:& |DS,q\rangle \langle 0| \rightarrow e^{i\theta_{\pi/3}^q} 
  |DS,q\rangle\langle 0|, 
\end{eqnarray} 
where the last line holds only in the isotropic limit. 
We will now examine the general constraints on the above eigenvalues.   
 
{\it Inversion}.  First, one can show that the phases $\theta_\pi^q$ must 
be independent of $q$ by using the commutation relation 
$[R_\pi, {\mathcal T}_{\rm ferm}] = 0$ when acting on 
\emph{half-filled} states, which are gauge invariant.  For example, it 
follows from $[R_\pi, {\mathcal T}_{\rm ferm}]F_{0,q}^\dagger 
|DS,q\rangle = 0$ that $\theta_\pi^q = \theta_\pi^{-q}$.  Furthermore, 
$R_\pi^2 = 1$ implies that $e^{i\theta_\pi^q} \equiv \zeta_{\pi} = 
\pm 1$, so we have  
\begin{equation} 
  R_\pi : |DS,q\rangle \langle 0|  
  \rightarrow \zeta_{\pi} |DS,q\rangle \langle 0|. 
\end{equation} 
The sign $\zeta_{\pi}$ is in principle fixed, but can 
not be determined using general relations alone.

{\it Modified reflection}.  Similarly, it follows from 
the commutation relations 
$[\tilde {\mathcal R}_x,{\mathcal C}]  
= [\tilde {\mathcal R}_x,{\mathcal T}_{\rm ferm}] =0$ (on physical 
states) that 
\begin{equation} 
  \tilde {\mathcal R}_x : |DS,q\rangle \langle 0| \rightarrow 
  i\zeta_x |DS,q\rangle \langle 0|. 
\end{equation} 
The sign $\zeta_x = \pm 1$ is also fixed, but  
can not be determined from this general analysis.

{\it Translations}.  We constrain the phases $\theta^q_{1,2}$ by first  
assuming the 
following operator relations hold when acting on gauge-invariant states, 
\begin{eqnarray} 
  T_2 R_\pi &=& R_\pi T_2^{-1}, \\ 
  T_2 \tilde {\mathcal R}_x &=& \tilde {\mathcal R}_x T_1 T_2 ~, 
\end{eqnarray} 
since the left and right sides transform the lattice identically.   
The first relation implies $e^{i\theta^q_2} = \pm 1$, while  
using the second we conclude $e^{i\theta^q_1} = -1$. 
We can fix the former sign by now specializing to the 
isotropic limit.  Here, we have an additional symmetry relation, 
\begin{equation} 
  T_1 R_{\pi/3} = R_{\pi/3} T_2^{-1}, 
\end{equation} 
that holds when acting on physical states.  From this we obtain  
$e^{i\theta^q_1} = e^{i\theta^q_2} = -1$ in 
the isotropic limit.  By continuity, we assume 
this carries over in the anisotropic limit as well so that 
\begin{equation} 
  T_{1,2} : |DS,q\rangle\langle 0| \rightarrow -|DS,q\rangle\langle 
  0|. 
\end{equation} 
 
{\it Rotations (isotropic limit)}.  The commutation relation 
$[R_{\pi/3},{\mathcal T}_{\rm ferm}] = 0$ 
on physical states implies that  
$e^{i\theta_{\pi/3}^{-q}} = e^{-i\theta_{\pi/3}^q}$. 
Moreover, the relation $R_{\pi/3}^3 = R_\pi$ together with commutation 
with particle-hole symmetry yields $e^{i\theta_{\pi/3}^q} = \zeta_{\pi} 
e^{2iq\pi/3}$.  Thus, we have 
\begin{equation} 
  R_{\pi/3} : |DS,q\rangle \langle 0| \rightarrow 
  \zeta_{\pi} e^{2iq\pi/3} |DS,q\rangle \langle 0|. 
\end{equation} 
 
We have now arrived at the transformation properties listed in Eq.\ 
(\ref{DiracSeaSymmetry}).  As discussed in Sec.\ \ref{S+-},  
determining the ambiguities in $\zeta_{\pi}$ and $\zeta_x$  
that arose above is crucial for 
understanding the in-plane spin correlations in the AVL.  We will 
attempt to sort out these uncertainties by appealing to numerics, 
discussed below. 
 
\subsubsection{Numerical diagonalization} 
 
For convenience, 
we specialize to the isotropic limit $J = J'$ for the remainder of 
this appendix.   
Consider the mean-field Hamiltonian (\ref{MeanFieldH}) generalized to 
include arbitrary flux insertions, 
\begin{equation} 
  {\mathcal H}_{\rm MF} = -t\sum_{\langle {\bf x x'} \rangle}  
  [d^\dagger_{\bf x} d_{\bf x'} e^{-i (a^0_{\bf x x'}+\delta 
  a^0_{\bf x x'})} + 
  \text{H.c.}]. 
\end{equation} 
Here $a^0$ is the gauge field giving rise to a background $\pi$ flux 
as before, while $\delta a$ gives rise to any added flux.     
We have numerically  
diagonalized the above Hamiltonian on finite systems to obtain the 
spectrum of single-particle energies and wave functions. 
Specifically, we considered lattices composed 
of $N_{\rm rings}$ concentric ``rings'' of honeycomb sites.   
For example, Figs.\ \ref{CDWs} and \ref{SuperSolid} illustrate systems 
with $N_{\rm rings} = 2$ and 3, respectively.  System sizes up to 
$N_{\rm rings} = 17$, consisting of 1734 lattice sites, were studied.   
Flux insertions varying from 0 to $2\pi$ were taken to be uniformly spread 
within the first several innermost rings.   
A variety of boundary  
conditions were used, namely, open boundary 
conditions; ``Klein bottle'' boundary conditions, where one connects  
boundary sites with coordination number 2  
and their inversion 
counterparts; and modifications of the latter, where one connects only 
a subset of such boundary sites and their inversions. 
(Note that open boundary conditions are problematic when 
$N_{\rm rings}$ is odd, because with no added flux there 
are two degenerate zero-energy modes and therefore no unique ground 
state $|0\rangle$.  Other boundary conditions mentioned above 
give a unique ground state as desired.)  Such geometries break 
translational invariance, but are particularly convenient for 
addressing the symmetries of interest here.   
 
\begin{figure}  
  \begin{center}  
    {\resizebox{7cm}{!}{\includegraphics{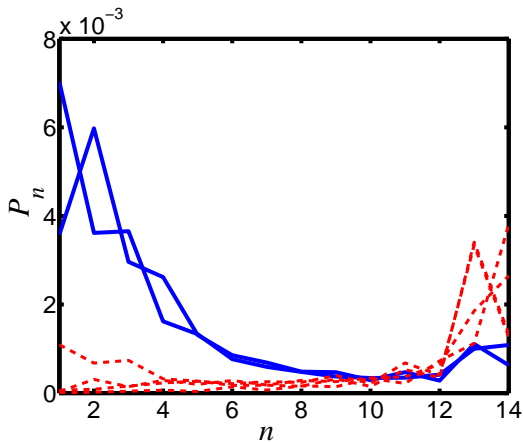}}}  
  \end{center}  
  \caption{Ring participation $P_n$ versus the ring index $n$ for the 
  first several wave functions above zero energy.  The data correspond 
  to a system with $N_{\rm rings} = 
  14$, open boundary conditions, and a $2\pi$ flux insertion spread 
  within the first four innermost rings.  The  quasi-localized modes 
  (solid curves), whose wave functions are peaked near the flux 
  insertion, are clearly distinguishable from other low  
  energy modes (dashed lines).   }  
  \label{RingParticipation}  
\end{figure}

With the single-particle wave functions in hand, one can 
explicitly construct the states $|0\rangle$ and $|DS,+1\rangle$.  The 
ground state $|0\rangle$ is simply built from all negative-energy 
wave functions.  More care must be taken, however, in constructing 
$|DS,+1\rangle$.  In a finite system, the four quasi-localized  
``zero-modes''---which are excluded from this state---are split away 
from zero energy, two above zero and two below. 
Identifying these quasi-localized modes is 
essential, particularly since the boundaries can introduce additional 
``edge'' modes 
near zero energy.  These modes can be distinguished by the behavior of 
their wave functions.  Most of the probability weight lies near the flux 
insertion for the quasi-localized modes, whereas the dominant weight for 
the edge modes occurs near the boundary.  A useful diagnostic  
for this comparison is the ``ring participation'' 
$P_n(\phi)$, which for a particular wave function $\phi$ gives  
the probability weight summed over honeycomb ring $n$, normalized by the number 
of sites in the ring.  More explicitly, 
\begin{equation} 
  P_n(\phi) = \frac{1}{N_{\rm sites}(n)}  
  \sum_{i\in n} |\phi(i)|^2, 
\end{equation} 
where $N_{\rm sites}(n)$ is the total number of sites in ring $n$ and 
$i$ is summed over all sites in the ring.  Figure \ref{RingParticipation} 
displays the ring participation for the first several wave functions 
above zero energy in a system with $N_{\rm rings} = 14$, 
open boundary conditions, and $2\pi$ flux inserted within the first 
four innermost rings. 
This illustrates the clear difference between the 
quasi-localized modes (solid lines) and other low-energy states 
(dashed lines).  In most cases observed this distinction  
allows one to identify the former, which are the modes of interest.   
Once these 
have been located, the state $|DS,+1\rangle$ can be built out of 
the remaining negative-energy wave functions.   
 
The inversion eigenvalues for $|0\rangle$ and $|DS,+1\rangle$ are then 
simply given by the product of inversion eigenvalues for the 
single-particle wave functions contributing to these states.   
We find that the parity under inversion for $|0\rangle$ and 
$|DS,+1\rangle$ depends on both the 
system size and boundary conditions.  However, in all cases where 
the quasi-localized modes can be clearly resolved, the \emph{product} 
of the inversion eigenvalues of $|0\rangle$ and  
$|DS,+1\rangle$---which gives the sign $\zeta_R$---is geometry 
independent.  In particular, we find $\zeta_R = -1$.   
 
Some insight into this result can be obtained  
by viewing the $2\pi$ flux as being inserted 
adiabatically.  Numerically, this is achieved by ramping the added 
flux from 0 to $2\pi$ in several stages and monitoring the evolution 
of the energy levels during the insertion.  In all cases studied, 
no levels cross zero energy during the evolution (though we do not 
have an argument for why this is the case).  This implies  
that the quantum numbers for the state with all  
negative-energy modes occupied are 
unchanged by the $2\pi$ flux insertion.  Moreover, we observe that  
the two quasi-localized modes 
with energy slightly below zero always have \emph{opposite} inversion 
eigenvalues.   Consequently, the states $|0\rangle$ and $|DS,+1\rangle$ 
must also have opposite inversion parity.  Why the quasi-localized  
modes split in this way is unclear at the moment, but would be 
useful to explore.   
 
Obtaining the sign $\zeta_x$ from numerics is more subtle due to the 
antiunitarity of $\tilde {\mathcal R}_x$.  A more useful symmetry 
to examine is the unitary operation ${\mathcal R}_x'  
= \tilde {\mathcal R}_x {\mathcal C} {\mathcal T}_{\rm ferm}$, which 
has eigenvalues $\pm 1$.  We will use the 
eigenvalues of ${\mathcal R}_x'$ to back out the sign $\zeta_x$.   
The fact that we find no zero-energy level crossings provides  
a useful shortcut to this end (but is not necessary). 
Again, the quantum numbers of the state with all negative-energy modes 
filled are then conserved under a $2\pi$ flux insertion.  In 
particular, the ground state 
$|0\rangle$ and the $q = 1$ Dirac sea with the two negative-energy  
quasi-localized modes filled must have identical  
eigenvalues under both ${\mathcal R}_x'$ and $R_{\pi/3}$.  There are 
just two candidates for the latter state, since only  
$(F_{R,+1}^\dagger -F_{L,+1}^\dagger)|DS,+1\rangle$ and 
$(F_{1,+1}^\dagger + F_{2,+1}^\dagger + 
F_{3,+1}^\dagger)|DS,+1\rangle$ have the same rotation  
eigenvalue as the ground state $|0\rangle$.  Furthermore,  
under ${\mathcal R}_x'$ both candidates have eigenvalues that differ 
from $|0\rangle$ by $-\zeta_x$.  Hence in either case we conclude that  
$\zeta_x = -1$.

\begin{table*} 
\caption{\label{tab:NonEnhanced} Transformation properties of the 48 bilinears 
which are components of conserved currents at the AVL fixed point and thus  
do \emph{not} have enhanced correlations.  In the first column 
$\sigma_\pm = \sigma^x \pm i \sigma^y$.}  
\begin{ruledtabular}  
\begin{tabular}{c | c | c | c | c | c | c}  
  & $T_{\delta{\bf r}}$ & $\tilde {\cal R}_x$ & $R_\pi$  
  & ${\mathcal C}$ & ${\cal T}_{\rm ferm}$ & $R_{\pi/3}$ (isotropic limit) 
  \\ \hline 
  $\psi^\dagger \psi = \rho \to$ & $\rho$ & $\rho$ & $\rho$ & $-\rho$  
  & $\rho$ & $\rho$  
  \\ \hline 
  $\psi^\dagger \sigma_{\pm} \psi = G_{\pm} \to$ & $G_{\pm}$  
  & $G_{\pm}$ & $-G_{\pm}$ & $-G_{\pm}$ & $-G_{\mp}$  
  & $e^{\pm i \frac{\pi}{3}}G_{\pm}$  
  \\ \hline 
  $\psi^\dagger \mu^z\tau^z \psi = M^{zz} \to$ & $M^{zz}$ & $M^{zz}$ 
  & $-M^{zz}$ & $M^{zz}$ & $-M^{zz}$ & $-M^{zz}$ 
  \\ \hline 
  $\psi^\dagger \mu^z\tau^z \sigma_{\pm} \psi = M^{zz}_{\pm} \to$  
  & $M^{zz}_{\pm}$ & $M^{zz}_\pm$ 
  & $M^{zz}_\pm$ & $M^{zz}_\pm$ & $M^{zz}_\mp$ & $-e^{\pm i 
  \frac{\pi}{3}} M^{zz}_\pm$ 
  \\ \hline 
  $\psi^\dagger (\tau^x + i \mu^z\tau^y) \psi = N \to$  
  & $e^{i{\bf Q}\cdot{\bf r}}N$ & $N$ & $N^\dagger$ & $N$  
  & $N^\dagger$ & $N^\dagger$ 
  \\ \hline 
  $\psi^\dagger (\tau^x + i \mu^z\tau^y)\sigma_{\pm} \psi = N_{\pm} \to$  
  & $e^{i{\bf Q}\cdot{\bf r}}N_{\pm}$ & $N_{\pm}$ 
  & $-N_{\mp}^\dagger$ & $N_{\pm}$ & $-N_{\pm}^\dagger$ & $e^{\pm i 
  \frac{\pi}{3}} N_{\mp}^\dagger$ 
  \\ \hline 
  $\psi^\dagger \mu^x\tau^x \psi = \tilde {\mathcal K}_{1} \to$  
  & $e^{i {\bf K}_1\cdot {\bf \delta r}} \tilde {\mathcal K}_1$  
  & $\tilde {\mathcal K}_2$ & 
  $\tilde {\mathcal K}_1$ & $\tilde {\mathcal K}_1$  
  & $-\tilde {\mathcal K}_1$ & $\tilde {\mathcal K}_3$ 
  \\ \hline 
  $\psi^\dagger \mu^y \tau^x \psi = \tilde {\mathcal K}_{2} \to$  
  & $e^{i {\bf K}_2\cdot {\bf \delta r}} \tilde {\mathcal K}_2$  
  & $\tilde {\mathcal K}_1$ & $\tilde {\mathcal K}_2$  
  & $\tilde {\mathcal K}_2$ & $-\tilde {\mathcal K}_2$  
  & $\tilde {\mathcal K}_1$ 
  \\ \hline 
  $\psi^\dagger \mu^z \psi = \tilde {\mathcal K}_{3} \to$  
  & $e^{i {\bf K}_3\cdot {\bf \delta r}} \tilde {\mathcal K}_3$  
  & $\tilde {\mathcal K}_3$ & $\tilde {\mathcal K}_3$  
  & $\tilde {\mathcal K}_3$ & $-\tilde {\mathcal K}_3$  
  & $\tilde {\mathcal K}_2$ 
  \\ \hline 
  $\psi^\dagger \mu^x\tau^x\sigma_\pm \psi = \tilde {\mathcal K}_{1\pm} \to$  
  & $e^{i {\bf K}_1\cdot {\bf \delta r}} \tilde {\mathcal K}_{1\pm}$  
  & $\tilde {\mathcal K}_{2\pm}$ & $-\tilde {\mathcal K}_{1\pm}$  
  & $\tilde {\mathcal K}_{1\pm}$  
  & $\tilde {\mathcal K}_{1\mp}$  
  & $e^{\pm i \frac{\pi}{3}} \tilde {\mathcal K}_{3\pm}$ 
  \\ \hline 
  $\psi^\dagger \mu^y \tau^x \sigma_\pm \psi = \tilde {\mathcal K}_{2\pm} \to$  
  & $e^{i {\bf K}_2\cdot {\bf \delta r}} \tilde {\mathcal K}_{2\pm}$  
  & $\tilde {\mathcal K}_{1\pm}$ & $-\tilde {\mathcal K}_{2\pm}$  
  & $\tilde {\mathcal K}_{2\pm}$ & $\tilde {\mathcal K}_{2\mp}$  
  & $e^{\pm i \frac{\pi}{3}} \tilde {\mathcal K}_{1\pm}$ 
  \\ \hline 
  $\psi^\dagger \mu^z \sigma_\pm \psi = \tilde {\mathcal K}_{3\pm} \to$  
  & $e^{i {\bf K}_3\cdot {\bf \delta r}} \tilde {\mathcal K}_{3\pm}$  
  & $\tilde {\mathcal K}_{3\pm}$ & $-\tilde {\mathcal K}_{3\pm}$  
  & $\tilde {\mathcal K}_{3\pm}$  
  & $\tilde {\mathcal K}_{3\mp}$  
  & $e^{\pm i \frac{\pi}{3}} \tilde {\mathcal K}_{2\pm}$ 
  \\ \hline 
  $\psi^\dagger \mu^y \tau^y \psi = \tilde {\mathcal K}'_{1} \to$  
  & $e^{i {\bf K}_1\cdot {\bf \delta r}} \tilde {\mathcal K}'_1$  
  & $-\tilde {\mathcal K}'_2$ & 
  $-\tilde {\mathcal K}'_1$ & $-\tilde {\mathcal K}'_1$  
  & $\tilde {\mathcal K}'_1$ & $\tilde {\mathcal K}'_3$ 
  \\ \hline 
  $\psi^\dagger \mu^x\tau^y \psi = \tilde {\mathcal K}'_{2} \to$  
  & $e^{i {\bf K}_2\cdot {\bf \delta r}} \tilde {\mathcal K}'_2$  
  & $-\tilde {\mathcal K}'_1$ & 
  $-\tilde {\mathcal K}'_2$ & $-\tilde {\mathcal K}'_2$  
  & $\tilde {\mathcal K}'_2$ & $\tilde {\mathcal K}'_1$ 
  \\ \hline 
  $\psi^\dagger \tau^z \psi = \tilde {\mathcal K}'_{3} \to$  
  & $e^{i {\bf K}_3\cdot {\bf \delta r}} \tilde {\mathcal K}'_3$  
  & $\tilde {\mathcal K}'_3$ & 
  $-\tilde {\mathcal K}'_3$ & $-\tilde {\mathcal K}'_3$  
  & $\tilde {\mathcal K}'_3$ & $-\tilde {\mathcal K}'_2$ 
  \\ \hline 
  $\psi^\dagger \mu^y \tau^y \sigma_\pm \psi=\tilde {\mathcal K}'_{1\pm} \to$  
  & $e^{i {\bf K}_1\cdot {\bf \delta r}} \tilde {\mathcal K}'_{1\pm}$  
  & $-\tilde {\mathcal K}'_{2\pm}$ & $\tilde {\mathcal K}'_{1\pm}$  
  & $-\tilde {\mathcal K}'_{1\pm}$  
  & $-\tilde {\mathcal K}'_{1\mp}$  
  & $e^{\pm i\frac{\pi}{3}} \tilde {\mathcal K}'_{3\pm}$ 
  \\ \hline 
  $\psi^\dagger \mu^x\tau^y \sigma_\pm \psi = \tilde {\mathcal K}'_{2\pm} \to$  
  & $e^{i {\bf K}_2\cdot {\bf \delta r}} \tilde {\mathcal K}'_{2\pm}$  
  & $-\tilde {\mathcal K}'_{1\pm}$ & $\tilde {\mathcal K}'_{2\pm}$  
  & $-\tilde {\mathcal K}'_{2\pm}$  
  & $-\tilde {\mathcal K}'_{2\mp}$  
  & $e^{\pm i\frac{\pi}{3}} \tilde {\mathcal K}'_{1\pm}$ 
  \\ \hline 
  $\psi^\dagger \tau^z\sigma_\pm \psi = \tilde {\mathcal K}'_{3\pm} \to$  
  & $e^{i {\bf K}_3\cdot {\bf \delta r}} \tilde {\mathcal K}'_{3\pm}$  
  & $\tilde {\mathcal K}'_{3\pm}$ & $\tilde {\mathcal K}'_{3\pm}$  
  & $-\tilde {\mathcal K}'_{3\pm}$  
  & $-\tilde {\mathcal K}'_{3\mp}$  
  & $-e^{\pm i\frac{\pi}{3}} \tilde {\mathcal K}'_{2\pm}$ 
  \\ \hline 
  $\psi^\dagger (\mu^x + i \mu^y \tau^z) \psi = \tilde {\mathcal P}_1 \to $  
  & $e^{i {\bf P}_1\cdot {\bf \delta r}} \tilde {\mathcal P}_1$  
  & $\tilde {\mathcal P}_2$ & 
  $\tilde {\mathcal P}_1^\dagger$ & $-\tilde {\mathcal P}_1$  
  & $-\tilde {\mathcal P}_1^\dagger$  
  & $\tilde {\mathcal P}_3^\dagger$ 
  \\ \hline 
  $\psi^\dagger (\mu^y - i \mu^x \tau^z) \psi = \tilde {\mathcal P}_2 \to $  
  & $e^{i {\bf P}_2\cdot {\bf \delta r}} \tilde {\mathcal P}_2$  
  & $\tilde {\mathcal P}_1$ & 
  $\tilde {\mathcal P}_2^\dagger$ & $-\tilde {\mathcal P}_2$  
  & $-\tilde {\mathcal P}_2^\dagger$  
  & $\tilde {\mathcal P}_1^\dagger$ 
  \\ \hline 
  $\psi^\dagger (\mu^z \tau^x + i \tau^y) \psi = \tilde {\mathcal P}_3 \to $  
  & $e^{i {\bf P}_3\cdot {\bf \delta r}} \tilde {\mathcal P}_3$  
  & $\tilde {\mathcal P}_3$ & 
  $\tilde {\mathcal P}_3^\dagger$ & $-\tilde {\mathcal P}_3$  
  & $-\tilde {\mathcal P}_3^\dagger$ & 
  $\tilde {\mathcal P}_2^\dagger$  
  \\ \hline 
  $\psi^\dagger (\mu^x + i \mu^y \tau^z)\sigma_\pm \psi = \tilde 
  {\mathcal P}_{1\pm} \to $  
  & $e^{i {\bf P}_1\cdot {\bf \delta r}} \tilde {\mathcal P}_{1\pm}$  
  & $\tilde {\mathcal P}_{2\pm}$ & $-\tilde {\mathcal P}_{1\pm}^\dagger$  
  & $-\tilde {\mathcal P}_{1\pm}$ & $\tilde {\mathcal P}_{1\pm}^\dagger$  
  & $e^{\pm i \frac{\pi}{3}} \tilde {\mathcal P}_{3\mp}^\dagger$ 
  \\ \hline 
  $\psi^\dagger (\mu^y - i \mu^x \tau^z)\sigma_\pm \psi = \tilde 
  {\mathcal P}_{2\pm} \to $  
  & $e^{i {\bf P}_2\cdot {\bf \delta r}} \tilde {\mathcal P}_{2\pm}$  
  & $\tilde {\mathcal P}_{1\pm}$ & $-\tilde {\mathcal P}_{2\pm}^\dagger$  
  & $-\tilde {\mathcal P}_{2\pm}$ & $\tilde {\mathcal P}_{2\pm}^\dagger$  
  & $e^{\pm i \frac{\pi}{3}} \tilde {\mathcal P}_{1\mp}^\dagger$ 
  \\ \hline 
  $\psi^\dagger (\mu^z \tau^x + i \tau^y)\sigma_\pm \psi = \tilde  
  {\mathcal P}_{3\pm} \to$  
  & $e^{i {\bf P}_3\cdot {\bf \delta r}} \tilde {\mathcal P}_{3\pm}$  
  & $\tilde {\mathcal P}_{3\pm}$ & $-\tilde {\mathcal P}_{3\pm}^\dagger$  
  & $-\tilde {\mathcal P}_{3\pm}$ & $\tilde {\mathcal P}_{3\pm}^\dagger$  
  & $e^{\pm i \frac{\pi}{3}} \tilde {\mathcal P}_{2\mp}^\dagger$  
\end{tabular}  
\end{ruledtabular}  
\end{table*}


\end{document}